\documentclass[aps,prl,twocolumn, showpacs,dvipsnames,superscriptaddress]{revtex4-1}

\usepackage{amssymb,amsfonts,amsmath} 
\usepackage{graphicx,epsfig,psfrag}
\usepackage{color}
\usepackage{natbib}
\usepackage{url}
\usepackage[breaklinks=true]{hyperref}
\usepackage{mathtools}
\usepackage{subfigure}
\hypersetup{
        colorlinks=true,
        citecolor    = blue
}
\usepackage{bookmark}
\usepackage{epic}
\usepackage{longtable}

\newcommand{\pdag}{\phantom\dag}
\usepackage{bbm}
\usepackage{ulem}
\usepackage{braket}
\usepackage{xcolor}
\usepackage{csquotes}
\usepackage{graphicx}
\usepackage{tikz}
\usepackage{extarrows}
\usetikzlibrary{trees}
\usetikzlibrary{decorations.pathmorphing}
\usetikzlibrary{decorations.markings}
\usetikzlibrary{positioning,arrows}
\usetikzlibrary{shapes.arrows}
\usepackage{leftidx}
\usepackage{framed}
\usepackage{tcolorbox}
\usepackage{bm}
\usepackage{MnSymbol}

\tikzset{->-/.style={decoration={
  markings,
  mark=at position #1 with {\arrow{>}}},postaction={decorate}}} 

\begin{document} 

\title{A Luttinger Liquid coupled to Ohmic-class  environments}

\author{Andisheh Khedri}
\affiliation{Institute  for  Theoretical  Physics,  ETH  Zurich,  8093  Zurich,  Switzerland}
\author{Antonio \v{S}trkalj}
\affiliation{Institute  for  Theoretical  Physics,  ETH  Zurich,  8093  Zurich,  Switzerland}
\author{Alessio Chiocchetta} 
\affiliation{Institute for Theoretical Physics, University of Cologne, D-50937 Cologne, Germany}
\author{Oded Zilberberg} 
\affiliation{Institute  for  Theoretical  Physics,  ETH  Zurich,  8093  Zurich,  Switzerland}
\begin{abstract} 

We investigate the impact of an Ohmic-class environment on the conduction and correlation properties of one-dimensional interacting systems. Interestingly, we reveal that inter-particle interactions can be engineered by the environment's noise statistics. 
Introducing a backscattering impurity to the system, we address Kane-Fisher's metal-to-insulator quantum phase transition in this  noisy and realistic setting.
Within a perturbative renormalization group approach, we show that the Ohmic environments keep the phase transition intact, while sub- and super-Ohmic environments, modify it into a smooth crossover at a scale that depends on the interaction strength within the wire. 
The system still undergoes a metal-to-insulator-like transition when moving from sub-Ohmic to super-Ohmic environment noise.
We cover a broad range of realistic experimental conditions, by exploring the impact of a  finite wire length and temperature on transport through the system. 

\end{abstract}
\date{\today} 
\maketitle

One of the most fascinating manifestations of quantum many-body physics
occurs in one-dimensional systems. There, irrespective of whether the interacting particles are gapless fermions, bosons or spins, their low-energy properties universally-exhibit  \textit{Tomonaga-Luttinger liquid} (TLL) behaviour with well-defined bosonic excitations 
~\cite{Delft98,Haldane81,Giamarchi04,gogolin2004,Cazalilla11}.
Distinct signatures of TLL~\cite{Giamarchi12}, include  separation of spin and charge degrees of freedom~\cite{Auslaender02,Jompol09,Bocquillon13,Hashisaka17}, which are experimentally verified as fractionalization of injected charges~\cite{Steinberg08,Prokudina14,Kamata14,Freulon15}; power-law behavior of correlation functions, also known as the \textit{zero-bias anomaly}~\cite{Matveev93,Bockrath99,Yao99}; and \textit{Kane-Fisher} impurity physics~\cite{Kane92a,Kane92b}.
The latter concerns the sensitivity of gapless excitations to local perturbations,
that are microscopically rationalized in terms of Friedel oscillations~\cite{Glazman93}, and as a manifestation of \textit{orthogonality catastrophe}~\cite{Fabrizio96}.
Thus, the presence of a backscattering impurity inside a TLL~\cite{Kane92a,Kane92b,Anthore18} engenders a quantum phase transition between a perfectly-conducting phase and an insulating phase 
as a function of interaction strength.
Such TLL features have been observed in a wide variety of experiments including nanotubes~\cite{Bockrath99,Yao99,Cao05}, quantum Hall edges~\cite{Wen90,Chang03}, cold-atom platforms~\cite{Yang17,Esslinger18,Yang18}, circuit quantum simulations~\cite{Anthore18},
antiferromagnetic spin chains~\cite{Lake2005}, and spin ladder systems~\cite{Dagotto1999,Giamarchi2008}.

Recent technological advances in solid-state platforms~\cite{Prokudina14,Fujisawa18,Yang2020}, as well as many-body quantum simulators with cold atom experiments~\cite{Bloch08,Bloch12,Nori14,Langen2015,Krinner17,Esslinger18}, 
or quantum circuits~\cite{Jezouin2013,Anthore18}, 
dynamics of open quantum systems~\cite{Eisert15}. 
Here, the competition between coherent quantum processes and incoherent forcing induced by the environment leads to novel physics, with no counterpart in isolated quantum systems~\cite{Neto97,Cazalilla06,Nattermann08,Altland12,Soriente_2018,Marino19,dogra2019dissipation,Soriente_2020,Ferguson_2020,Soriente2021}, and raises
fundamental question regarding the existence of universality in open quantum systems.
As a result, the interplay between TLL physics, dissipation, and drive is also revisited, leading to novel effects, such as  many-body quantum Zeno effect due to localized loss~\cite{Barontini13,Labouvie16,Froeml19a,Froeml19b,Zezyulin12,Bu2020}, engineering correlation with non-local two-body loss effects \cite{Syassen2008,Cirac2009}, exotic phases such as Zeno insulators, and dissipation-induced spin-charge separation~\cite{Ueda20}.
Similarly, the impact of out-of-equilibrium scenarios have been extensively explored~\cite{Gutman08,Gutman09,Gutman10,Ngo10,Marino19,Karrasch12} highlighting further universal open TLL physics. 

A particularly-relevant instance of an open TLL involves the presence of leads (reservoirs) attached to the ends of a quantum wire. This setup is fundamental for exploring quantum transport
~\cite{Tarucha95,Ogata94,Krinner17,Esslinger18}.
However, the presence of leads (and their resulting dissipative channels) unavoidably affect the wire's transport properties,  
where even the presence of non-interacting Ohmic leads melts the insulating phase of dirty TLLs~\cite{Furusaki96,Maslov95,Furusaki93,Kane93,Maslov95,Ngo10}, as well as the aforementioned zero-bias anomaly~\cite{Antonio19}.
Realistic leads, however, are realized by distinctly-different systems, ranging from metallic gates with an unscreened Coulomb potential~\cite{Fabrizio1994} or lattice vibrations~\cite{Sethna82}, to superfluid ultracold gases~\cite{husmann2015connecting}, or to complex RC circuits~\cite{Weiss99}. Accordingly, these realizations bring about different open system scenarios, with various environment densities of states, coupling to the TLL, and fluctuations. This diversity can result in vastly different phenomena. Despite its crucial role, to our knowledge, a unifying, low-energy theory that is able to include the effect of arbitrary reservoirs coupled to a TLL is still missing.

In this work, we fill this gap by formulating a low-energy theory for a quantum wire in contact with arbitrary Ohmic-class leads, and show how these dramatically modify the low-energy properties of the wire.
To highlight the consequences of this open system interplay, we analyze Kane-Fisher's impurity problem as a concrete setup. 
We show that while a super-Ohmic environment (with fast fluctuations)
localizes the particles akin to a Zeno effect, 
sub-Ohmic noise (slow fluctuations)
overwhelm 
the low-energy properties of the TLL. The impurity, then, engenders the  TLL--environment competition with a non-monotonous renormalization group (RG) flow, leading ultimately to a conducting-to-insulating-like transition in the TLL as a function of the noise statistics in the environment. Furthermore, considering realistic finite-size 1D systems, we predict that the non-monotonous flow implies unusual temperature-dependent scaling of the conductance coming out from the TLL--environment competition.

\paragraph*{Setup and microscopic model ---}
%
\begin{figure}
    \centering
    \includegraphics[scale=1]{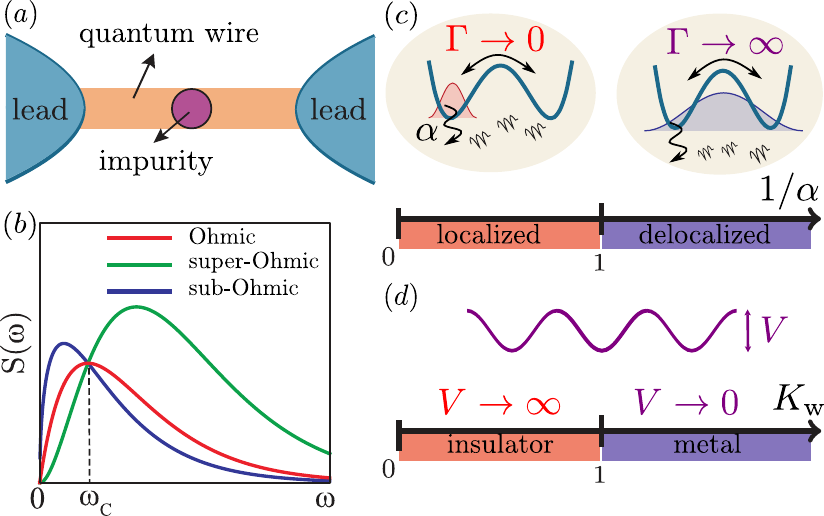}
    \caption{ 
    (a) 
    Sketch of a quantum wire containing a single impurity, that back-scatters electrons, coupled to leads.
        (b) 
    The noise-power spectrum of the electronic leads as a function of frequency, where $\omega_c$ marks the environment bandwidth. 
        (c)
    The phase diagram of a two-level system ($\phi^4$-theory) coupled to an Ohmic environment for varying dissipation strength $\alpha$. For small dissipation $\alpha<1$, the effective tunneling, $\Gamma$, between two potential wells diverges, i.e., the particle is delocalized, but for stronger dissipation $\alpha>1$ the particle is localized $\Gamma\to 0$~\cite{Leggett87}.
    (d) Phase diagram of the TLL hosting an impurity coupled to Ohmic leads [cf.~action~\eqref{eq:action}, with effective scattering potential $V$ mapped to a Sine-Gordon potential] as a function of the interaction strength in the wire $K_{\rm w}$ [cf.~Eq.~\eqref{eq:action}].
    }
    \label{fig:setup}
\end{figure}

We consider a system of interacting spinless electrons confined in a single-channel 1D wire of length $L$ that is adiabatically connected to metallic leads, see Fig.~\ref{fig:setup}(a).
The Hamiltonian of the wire reads
\begin{equation}
\resizebox{.91\hsize}{!}{$\displaystyle{
H^{\pdag}_{\rm w}=\int_x\, \bigg[
\sum_{\eta=\rm L,R}\left\{i\alpha_{\eta}\nu_{\rm F}\Psi^{\dag}_{\eta}(x)
\partial^{\pdag}_x \Psi^{\pdag}_{\eta}(x) 
+U\rho^{\pdag}_{\eta}(x)\rho^{\pdag}_{\eta}(x)\right\}\bigg]
\,,}$}
\label{eq:Hamiltonian}
\end{equation}
with $\int_x=\int_{-L/2}^{L/2} dx$,
where the first term represents the kinetic energy of electrons with linearized dispersion $\epsilon_k=\alpha_\eta  v^{\pdag}_{\rm F} k$,  $ v^{\pdag}_{\rm F}$ the electron velocity, and $\alpha^{\pdag}_{\rm L}= 1$ ($\alpha^{\pdag}_{\rm R}=-1$) corresponds to the left- (right-)moving electrons with fermionic field operators $\Psi^{\pdag}_{\rm L}$ ($\Psi^{\pdag}_{\rm R}$).
The second term describes local electron-electron interactions inside the wire via (normal-ordered) density operators $\rho^{\pdag}_{\eta}(x)= :\Psi_{\eta}^\dagger(x) \Psi^{\pdag}_{\eta}(x):$ with a constant magnitude $U$. 

Many-body interactions in the wire modify the relevant quasiparticles profoundly, resulting in the emergence of collective bosonic excitations~\cite{Delft98,Giamarchi04,gogolin2004}.
Using bosonization, we can write the fermionic fields as $\Psi^{\pdag}_{\eta} = \sqrt{\Lambda/(2\pi)} \hat{F}^{\pdag}_{\eta}\exp[ik^{\pdag}_{\rm F}x+i\phi^{\pdag}_{\eta}(x)]$, where $\Lambda$ is an ultraviolet cutoff, $\hat{F}^{\pdag}_{\eta}$ the Klein factor, and $\phi^{\pdag}_{\eta}$ represents bosonic fields with commutation relations,
$[\phi^{\pdag}_\eta (x),\phi^{\pdag}_{\eta^\prime}(y)]=-i \pi\,\,\alpha^{\pdag}_\eta\delta^{\pdag}_{\eta,\eta^\prime}  \text{sgn}(x-y)$. The density operator in terms of these bosonic fields reads $\rho^{\pdag}_{\eta}(x)=\partial^{\pdag}_x \phi^{\pdag}_{\eta}(x) / (2\pi)$. Thereby, the bosonized Hamiltonian of the interacting wire takes the form
$H^{\pdag}_{\rm w}=\dfrac{ v^{\pdag}}{4\pi}\int_x \left\{\dfrac{1}{K_{\rm w}}\left(\partial^{\pdag}_x\varphi\right)^2
+K_{\rm w}\left(\partial^{\pdag}_x\theta\right)^2\right\}$,
where $\varphi(x,t),\theta (x,t)=(1/\sqrt{2})\left[\phi^{\pdag}_{\rm L}(x,t)\pm \phi^{\pdag}_{\rm R}(x,t)\right]$ satisfy the commutation relation $[\varphi(x),\theta(y)]=i\pi\,\,\mathrm{sgn}(x-y)$, $\nu=\nu_{\rm F}/K_{\rm w}$, and $K_{\rm w}=1/\sqrt{1+U/(\pi v^{\pdag}_{\rm F})}$ is the so-called Luttinger Liquid parameter, with $K_{\rm w}=1$ referring to a noninteracting wire, and $K_{\rm w}<1$ ($K_{\rm w}>1$) indicating repulsive (attractive) interactions. 

The wire is connected to electronic leads, which we introduce by imposing appropriate boundary conditions $\partial^{\pdag}_t \phi^{\pdag}_{\eta}(x=\pm L/2)=2\pi J_{\eta}$,
where $J_{\eta}(\omega)$ is the current operator in the leads. 
The effect of the boundaries enter the correlation functions of the wire via the noise power spectrum $S(\omega)=\langle J_\eta(\omega) J_{\eta}(-\omega) \rangle$, where $\langle \cdots \rangle$ denotes thermal averaging with respect to the leads~\cite{Leggett87,Buttiker1992,Nazarov97,Antonio19}.
We consider an Ohmic-class noise power spectrum
\begin{equation}
S(\omega)=\omega \left|\frac{\omega}{\omega_c}\right|^{s-1}~ e^{-|\omega/\omega_c|} ~[1+n_b(\beta \omega)]\,,
\label{eq:noise_power_spectrum}
\end{equation}
where $\omega_c$ is the characteristic energy scale of the environment, indicating the exponential suppression of  current-current correlations for $\omega\gg\omega_c$. The parameter $s\in(0,2)$ distinguishes between different cases, i.e., $s=1$ describes an Ohmic lead, whereas $s < 1$ ($s>1$) corresponds to the sub- (super-)Ohmic case. The noise power exhibits a bosonic distribution
$n^{\pdag}_b(\beta\omega)=1/\left[\exp(\beta\omega)-1\right]$ at inverse temperature $\beta$. 

To realize an Ohmic environment, it suffices to consider free fermions with a well-defined Fermi-Dirac distribution. On the other hand, non-Ohmic environments with $s\neq 1$ can be realized, for example, by  electron-phonon coupling in the leads ($s>1$)~\cite{Sethna82}, or by complex RC circuit architectures ($s<1$)~\cite{Weiss99}.
In Fig.~\ref{fig:setup}(b), we plot the frequency dependence of the noise spectrum for these three cases. 
Comparing to the Ohmic case, the current-current fluctuations in the sub- (super)-Ohmic leads are more dominant at lower (higher) frequencies, i.e., environmental fluctuations are slower (faster).  
Ohmic-class environments have been extensively studied in the framework of the spin-boson model
\cite{Leggett87,Vojta05,Anders07,Weiss99,Kehrein95}, revealing the profound influence of the environment fluctuations on the nature of the ground state, as well as on the dynamics of the system. 
In particular, it was shown that in the Ohmic case, there exist a critical dissipation that distinguishes between a localized phase and a delocalized one, see Fig.~\ref{fig:setup}(c).
In contrast, in the sub- (super)-Ohmic case, the system is argued to be localized (delocalized) independent of dissipation strength \cite{Leggett87}.
Analogously, in this work, we investigate the impact of such current fluctuations in the leads on the transport through a disordered interacting wire.  

\paragraph*{Environment-induced correlations --- } 
First, we consider the limit of $T=0$, and investigate the impact of the noise-spectrum in the leads on the TLL physics. 
The correlations between bosonic excitations of TLL 
are entirely determined by the Hamiltonian~\eqref{eq:Hamiltonian} and the noise power spectrum at the boundaries~\eqref{eq:noise_power_spectrum}~\cite{Nazarov97,Antonio19}. In particular, the wire's bosonic greater Green's function defined as $\mathcal{G}^{>,0}_{\varphi\varphi}(x,x^\prime,\omega)\equiv - i\langle \varphi(x,\omega)\varphi^{\dagger}(x^\prime,\omega)\rangle_{0}$,
with $\langle \cdots \rangle_{0}$ referring here to the thermal average, is found to be
$\mathcal{G}^{>,0}_{\varphi\varphi}(x,x^\prime,\omega)=-iS(\omega)F_{\varphi}(x,x^\prime,\omega)/\omega^2$, with 
\begin{equation}
\resizebox{1.03\hsize}{!}{$\displaystyle{
F_{\varphi}(x,x^\prime,\omega)=
2 \frac{\left(\frac{1}{K_{\rm w}^2}-1\right)\cos[\frac{\omega\tau_L(x+x^\prime)}{L}]\cos[\omega\tau_L]
+\left(\frac{1}{K_{\rm w}^2}+1\right)\cos[\frac{\omega\tau_L(x-x^\prime)}{L}]}{\left(1+\frac{1}{K_{\rm w}^2}\right)^2-\left(1-\frac{1}{K_{\rm w}^2}\right)^2 \cos^2[\omega\tau_L]}\,,
}$}
\label{eq:structure}
\end{equation}
the structure function of a many-body Fabry-P\'{e}rot interferometer
that is formed due to the presence of the leads reflecting the bosonic excitations at the boundaries \cite{Nazarov97,Antonio19}.
Detailed-balance holds $\mathcal{G}^{<,0}_{\varphi\varphi}(x,x^\prime,\omega)\equiv - i\langle \varphi^{\dagger}(x^\prime,\omega)\varphi(x,\omega)\rangle_{0}=e^{-\beta\omega}\mathcal{G}^{>,0}_{\varphi\varphi}(x,x^\prime,\omega)$ as expected for bosons in thermal equilibrium~\cite{bruus04}.
The finite length of the wire introduces a characteristic time scale to the system, namely, the time of flight for the collective excitations $\tau_{\rm L}=LK_{\rm w}/ v^{\pdag}_{\rm F}$ to cross the wire. 
At high frequencies, $\omega\tau_{\rm L}\gg 1$, 
the system acts similarly to the infinite wire~\cite{supmat}, whereas at small frequencies,
$\omega\tau_{\rm L}\ll 1$, $F_{\varphi}(x,x^\prime,\omega)\approx1$ such that the physics of the interacting wire is washed out, and the system response is dominated by the environment.

\textit{An impurity in the wire at zero temperature --- } 
To reveal the consequences of the environment-induced correlations,
we consider a back-scattering impurity at $x=x_0$, leading to an additional  Hamiltonian term, $H_b= V^{\pdag}_{0}\left[\Psi^\dag_{\rm L}(x_0)\Psi^{\pdag}_{\rm R}(x_0)+\text{H.c.}\right]$.
The action of the bosonized system at all positions $x \neq x_0$ is quadratic and can be therefore integrated out, resulting in the following (\textit{local Sine-Gordon}) action in imaginary-time path-integral formalism
\begin{equation}
\resizebox{.91\hsize}{!}{$\displaystyle{\mathcal{A}=\int_{0}^{\beta}d\tau~\varphi^\dagger(\tau)
\left[ \mathcal{G}^{0}_{\varphi\varphi}(\tau) \right]^{-1} \varphi(\tau)
+V_0\int_{0}^{\beta}d\tau \cos[\sqrt{4\pi}\varphi(\tau)]\,,}$}
\label{eq:action}
\end{equation}
where $\mathcal{G}^0_{\varphi\varphi}(\tau)=\int_{-\infty}^{\infty} \mathrm{d} \omega \,  \mathcal{G}^0_{\varphi\varphi}(i\omega) e^{i\omega\tau}$, with $\mathcal{G}^0_{\varphi\varphi}(i\omega)=\int \frac{d\omega^{\prime}}{2\pi i} \frac{\mathcal{G}_{\varphi\varphi}^{0,>}(\omega^\prime)-\mathcal{G}_{\varphi\varphi}^{0,<}(\omega^\prime)}{i\omega-\omega^{\prime}}$  the imaginary-time (Matsubara) Green's function of the clean wire at $x=x_0$. 
Without loss of generality, we assume $x_0=0$~\footnote{The generalisation to $x_0 \neq 0$ can be readily carried out in an analogous ways, see for example Ref.~\cite{Furusaki96}.}. The action~\eqref{eq:action} and its corresponding correlation functions describe physics found in a variety of systems, including Brownian motion of a quantum mechanical particle in a periodic potential~\cite{Fisher85},
as well as in the dissipative two-level system ~\cite{Florens2010,Leggett87,Vojta05,Vojta12};
our specific wire--environment competition manifests through the explicit functional form of $\mathcal{G}^0_{\varphi\varphi}$.

The noise spectrum of Ohmic leads scales linearly at low energies, $S(\omega)\sim \omega$, and we commonly observe $\mathcal{G}^0_{\varphi\varphi}(k,i\omega)=K_{\rm w}/|2\omega|$. More generally, however, we can always define a similar structure $\mathcal{G}^0_{\varphi\varphi}(i\omega)=K(\omega)/|2\omega|$ with an energy-dependent Luttinger parameter $K(\omega)$ incorporating the finite length of the wire, and the fluctuations from the leads. At low frequencies $\omega\ll\omega_c$, we have~\cite{supmat}
\begin{equation}
\resizebox{.91\hsize}{!}{$\displaystyle{K(\omega)\approx\frac{K_{\rm w}}{\sin{(\pi s)/2}} \left|\frac{\omega}{\omega_c}\right|^{s-1}
\frac{1+K_{\rm w}+(1-K_{\rm w})e^{-\tau_{\rm L}|\omega|}}{1+K_{\rm w}-(1-K_{\rm w})e^{-\tau_{\rm L}|\omega|}}\,,}$}
\label{eq:D_mat}
\end{equation}
which is plotted in Fig.~\ref{fig:RG_flow}(a) for the long-wire limit $\omega \gg 1/\tau_L$. 
Equation~\eqref{eq:D_mat}, is one of the main results of this work, showcasing how the presence of leads, at the wire's ends, modifies the effective strength of the inter-particle interactions in the wire at different frequencies.
In particular, in the sub-Ohmic case (slow environmental fluctuations), at sufficiently low-frequencies, the effective interactions in the wire appear attractive [$K(\omega)>1$], while for the super-Ohmic case, the interactions become effectively repulsive [$K(\omega)<1$].

We employ a perturbative RG approach (due to the presence of infrared divergences \cite{Weiss99}), in which we integrate out high energy fields and map the system~\eqref{eq:action} to itself, but with a smaller ultraviolet cutoff $\Lambda^\prime=\Lambda(1-dl)$, i.e., $dl=d\Lambda/\Lambda$~\cite{Kane92a}. As a result, a renormalized scattering potential $V(\Lambda)$ (up to $dl^2$) obeys the flow equation
\begin{equation}
\frac{dV}{dl}=V_0 \left(1-K(\Lambda) \right) \, .  
\label{eq:RG_eq}
\end{equation}
\begin{figure}
    \centering
    \includegraphics[width=\columnwidth]{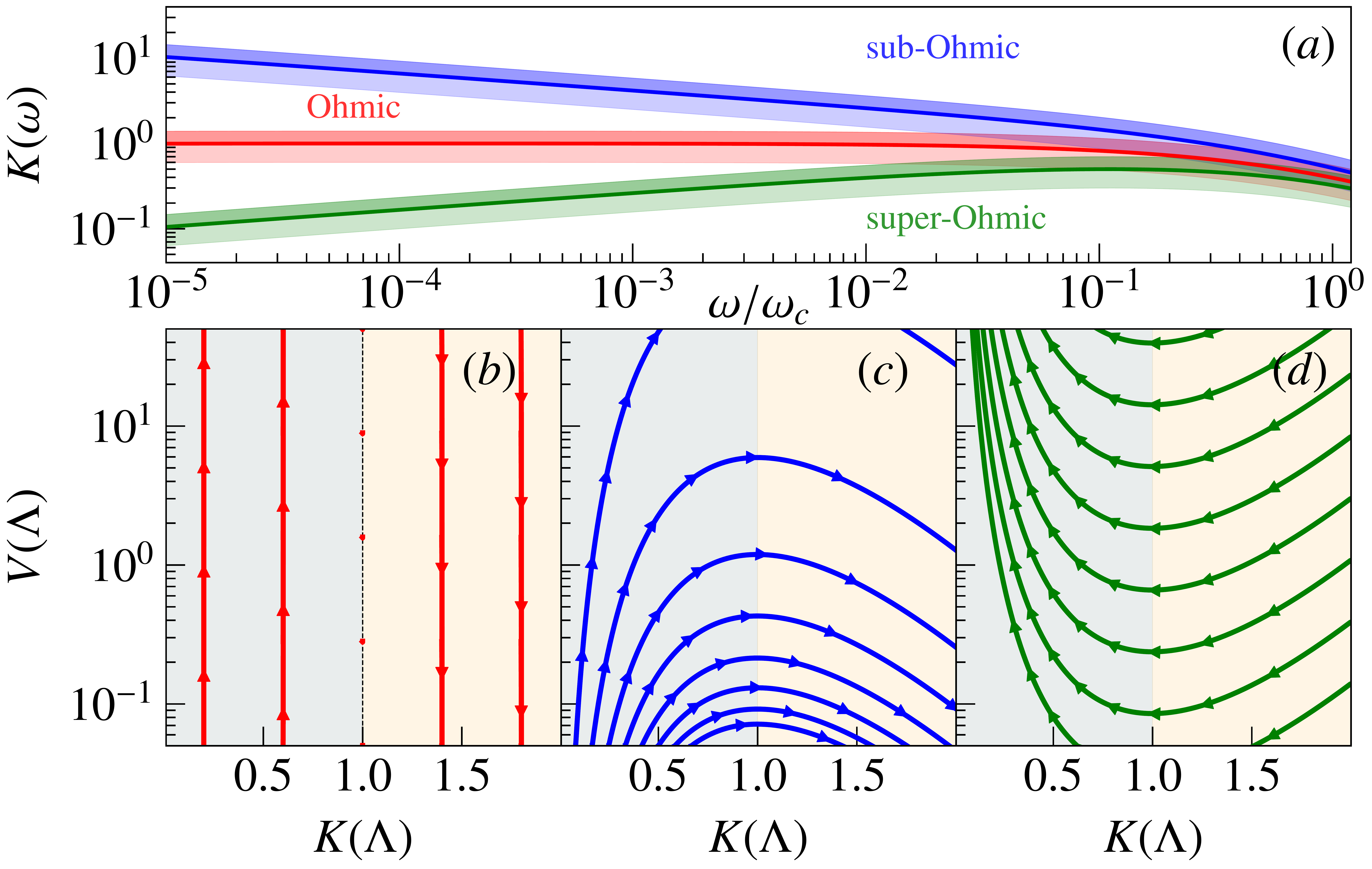}
    \caption{ 
   Scaling and RG-flow of the scattering potential $V(\omega)$ and effective Luttinger parameter $K(\omega)$
   at zero temperature, cf.~Eqs.~\eqref{eq:action}--\eqref{eq:RG_eq}. (a) The frequency-dependence of $K(\omega)$ for $\omega\gg 1/\tau_L$ for a range of interaction strengths in the wire $K_{\rm w}\in[0.6,1.4]$ and different Ohmic-class cases, $s=1.2,1.0,0.8$. The solid lines depict the non-interacting case $K_{\rm w}=1$, and the lighter (darker) shades mark $K_{\rm w}<1$ ($K_{\rm w}>1$).
   (b)-(d) The flow diagram (along the direction of the arrows) of the scattering potential versus effective interaction as we change the cutoff energy scale $\Lambda$ from $[\infty,\Lambda_{f}]$, with $\Lambda_{f}> 1/\tau_L$ for (b) Ohmic ($s=1.0$), (c) sub-Ohmic ($s=0.8$), and (d) super-Ohmic ($s=1.2$) cases.
    }
    \label{fig:RG_flow}
\end{figure}
Note that due to the noisy leads, the flow involves also the renormalization of $K(\omega)$. The numerical solution of the flow equation for the infinite wire limit is shown in Fig.~\ref{fig:RG_flow}, whereas impact of the finite wire is detailed in the Supplemental Material~\cite{supmat}. For the Ohmic case $s=1$ [Fig.~\ref{fig:RG_flow}(b)], standard \textit{Kane-Fisher physics}~\cite{Kane92a} is observed, where the wire's Luttinger parameter plays a decisive role, i.e., for $K_{\rm w}<1$ ($K_{\rm w}>1$) the fixed point of the RG, for $\Lambda\to 0$, is $V\to \infty$ ($V \to 0$)  for the insulating (metallic) case. The quantum critical point is at $K_{\rm w}=1$, corresponding to non-interacting electrons, see Fig.~\ref{fig:setup}(d). 

Considering sub-Ohmic environment noise, see Fig.~\ref{fig:RG_flow}(c), the low-frequency noise induces an effective $K(\omega)$ that increases with  $\Lambda$. Therefore, starting from repulsive interactions in the wire, the renormalized scattering potential exhibits a non-monotonic behaviour. Specifically, defining $\Lambda^*$ such that $K(\Lambda^*)=1$, we observe that for $\Lambda>\Lambda^*$, the back-scattering potential increases and transport through the wire is suppressed, while for $\Lambda<\Lambda^*$, the potential decreases and transport is unaffected by the impurity. The transition point $\Lambda^*$ strongly depends on the wire's bare Luttinger Liquid parameter $K_{\rm w}$, cf.~Eq.~\eqref{eq:D_mat}. Crucially, regardless of the specific initial microscopic parameters of the wire, the fixed point of the flow is $K\to \infty,V \to 0$. We compare the sub-Ohmic case with the super-Ohmic one, where the flow lines show an opposite trend, see Fig.~\ref{fig:RG_flow}(d). In this case, the effective $K(\omega)$ is reduced and the fixed point is realized at $K \to 0, V \to \infty$. 

Our perturbative RG analysis assumes an initially small scattering impurity potential. In the opposite limit of a strong impurity, we can formulate the problem in a dual representation with $\mathcal{G}_{\theta\theta}(i\omega)=1/(2\tilde{K}(\omega)|\omega|)$~\cite{supmat}, with $\tilde{K}(\omega)$ defined by changing $(1-s)\to (s-1)$  in Eq.~\eqref{eq:D_mat}. Hence, the sub-Ohmic environment acts as super-Ohmic one, and vice-versa~\cite{Weiss1996}. As a result, tunneling, $t$, across the barrier satisfies the RG equation $dt/dl=t_0(1-[1/\tilde{K}(\Lambda)])$, where $t_0$ is the initial tunneling amplitude.  We conclude that the metal-to-insulator quantum phase transition that occurs for the Ohmic environment at the critical point, $K_{\rm w}=1$, is replaced for non-Ohmic environments by a smooth cross-over with characteristic energy scale $\Lambda^*$ that depends on the interaction strength in the wire. 
\paragraph*{Conductance at finite temperatures --- }
We study ac-conductance through the wire, and consider an external probe in the form of a potential $U(x,t)=U(x)\cos(\omega t)$, leading to an additional term in the Hamiltonian, $\delta H=\sum_{\eta} \int \mathrm{d}x \rho_{\eta}(x) U(x)$.
Within linear response theory  \cite{Maslov95,Furusaki96,Glazman1997,supmat}, the ac-conductance reads 
$G(\omega)=-(e^2/h)2i\omega \mathcal{G}^{\rm R}_{\varphi\varphi}(\omega)$,
with the retarded plasmonic Green's function in the presence of the impurity  $\mathcal{G}^{\rm R}_{\varphi\varphi}(\omega)=\int \mathcal{D}[\varphi] \varphi(\omega)\varphi^{\dagger}(\omega)e^{-\mathcal{A}}|_{\omega+i0^+}=\left[(\mathcal{G}^{0,\rm R}_{\varphi\varphi})^{-1}(\omega)-\Sigma^{\rm R}(\omega)\right]^{-1}$, where the self-energy can be obtained by expanding the partition function corresponding to the action ~\eqref{eq:action}
~\cite{supmat}, which up to $V_0^2$ reads

\begin{equation}
\Sigma(i\omega)=i V_0^2\int_{0}^{\beta}d\tau~[1-e^{i\omega\tau}]
e^{E(\tau)},
\label{eq:self_energy}    
\end{equation}
with
\begin{equation}
E(\tau)=\int_{0}^{\infty}\frac{d\omega}{2\pi} 
\Bigg\{\frac{1-\cosh(\omega \tau)}{\tanh(\beta\omega/2)}
+\sinh(\omega \tau)\Bigg\} ~\mathcal{G}^0_{\varphi\varphi}(\omega).\
\end{equation}
Thereby, the ac-conductance can be written as $G(\omega)=(e^2/h)\left[G_0 - G_b\right]$,
where $G_0=-2i\omega \mathcal{G}^{\rm R,0}_{\varphi\varphi}(\omega)$ corresponds to the conductance through a clean wire, and $G_b \equiv -2i\omega \mathcal{G}^{\rm R,0}_{\varphi\varphi} \Sigma^{\rm R}(\omega)\mathcal{G}^{\rm R,0}_{\varphi\varphi}$ represents the correction to the conductance due to the presence of the impurity. 
In the dc-limit, $\lim_{\omega\to 0}G_0(\omega)=(\omega/\omega_c)^{s-1}$, which is independent of temperature and interaction strength inside the wire. For the Ohmic case, $\lim_{\omega\to 0} G_0(\omega)=1$, for sub-Ohmic this limit diverges, and in the super-Ohmic case it vanishes. 

\begin{figure}[t!]
    \centering
    \includegraphics[width=\columnwidth]{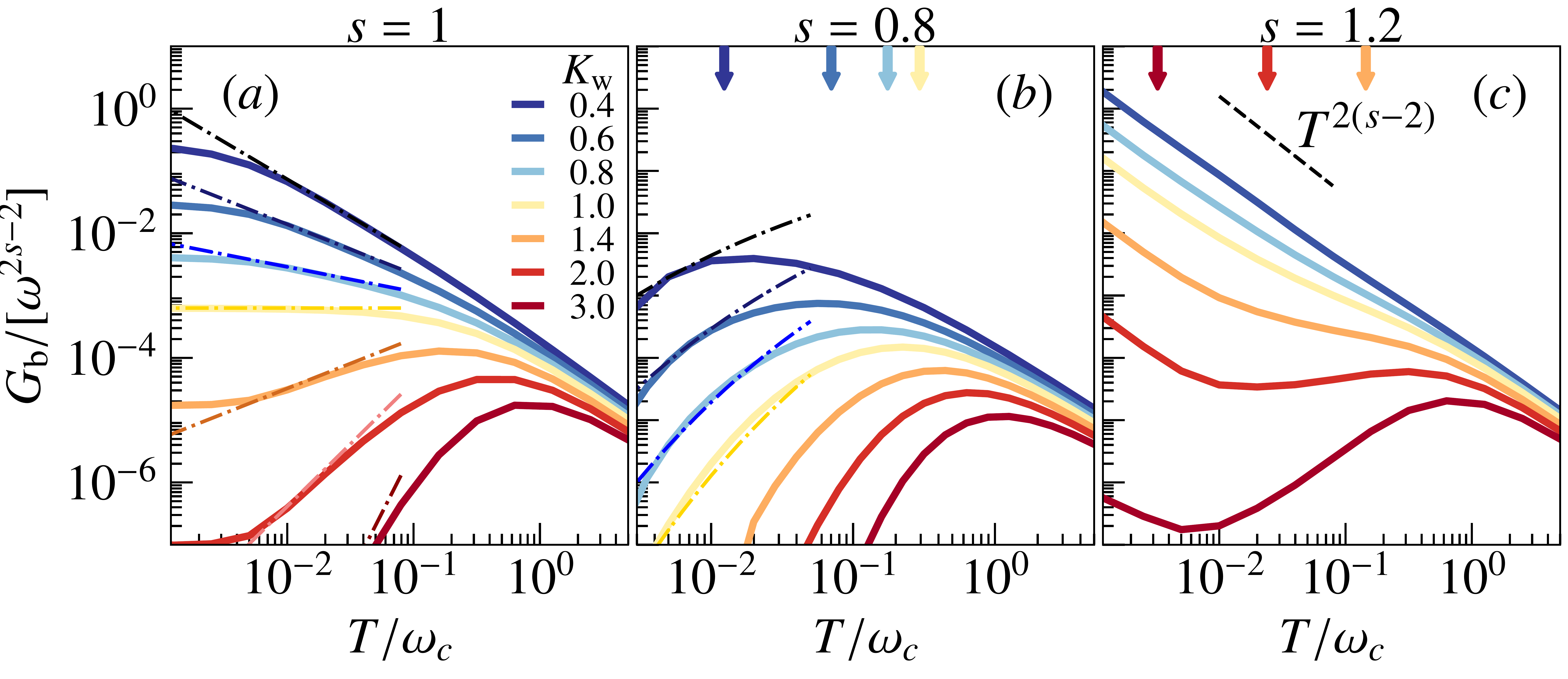}
    \caption{ 
     Temperature dependence of the impurity-induced correction to conductance through a finite-length wire [$L=100 \nu_{\rm F}/(K_{\rm w}\omega_{\rm c})$] for different interaction strengths ($K_{\rm w}$), cf.~Eq.~\eqref{eq:self_energy}. (a) Ohmic, (b) sub-Ohmic ($s=0.8$), and (c) super-Ohmic case ($s=1.2$). The dashed lines show in (a) the expected power-law dependence $G_b\propto (T/\omega_c)^{2K_{\rm w}-2}$, (b) the exponential dependence $G_b\propto \exp[-\alpha_s (T/\omega_c)^{s-1}]$, and (c) $G_b~\propto (T/\omega_c)^{2s-4}$~\cite{supmat}. The arrows in (b) and (c) marks $\Lambda^*$, where $K(\Lambda^*)=1$. 
   }
    \label{fig:conductance}
\end{figure}

In the following, we focus on the temperature-dependence of the correction  $G_b$, see Fig.~\ref{fig:conductance}. The temperature mimics the RG flow of the renormalized scattering potential (cf.~Fig.~\ref{fig:RG_flow}): 
(i) In the Ohmic case [Fig.~\ref{fig:conductance}(a)], for $1/\tau_{\rm L} \ll T \ll \omega_c$, we obtain a power-law temperature dependence of the form $G_b \propto (T/\omega_c)^{2K_{\rm w}-2}$. Such a power-law is characteristic for critical scaling close to a quantum phase transition~\cite{supmat}. Specifically, for repulsive interaction, $G_{\rm b}$ grows with decreasing temperature, while it gets suppressed for attractive interactions. For a non-interacting wire, $G_{\rm b}$ is independent of temperature. Interestingly, at small temperatures ($T\tau_{\rm L}\ll1$), we observe a temperature-independent behaviour for all values of $K_{\rm w}$, corresponding to the  cutoff of the critical scaling by the finite length of the wire; (ii)   
For the sub- and super-Ohmic case [Figs.~\ref{fig:conductance}(b) and (c), respectively], we observe a characteristic energy-scale $\Lambda^*$ above which the system qualitatively behaves as in the Ohmic case, i.e., with a power-law decrease governed by the noise scaling with $\omega_c$. However, at $T\ll \Lambda^*$, for the sub-Ohmic case, we obtain an exponential suppression with exponents depending on the interaction strength as $G_b\propto \exp[- K_{\rm w}\alpha_s (T/\omega_c)^{s-1}]$. In contrast, in the super-Ohmic case for $T\ll \Lambda^*$, the $G_{\rm b}$ grows with decreasing temperature in a power-law fashion with an exponent that is entirely independent of the interaction strength in the wire $T^{2s-4}$. The finite length effect at low temperatures is washed away by the noisy environment.

\paragraph*{Conclusion ---}
We show that the specifics of charge fluctuation at the boundaries of an interacting wire can modify the transport through a dirty Luttinger liquid beyond the Kane-Fisher description. 
We further outline the physical implications of the wire--environment competition for realistic transport measurements in a wide variety of systems.
Specifically, at low temperatures, the impurity-induced correction to conductance (i) follows the result of Kane-Fisher~\cite{Kane92a}, and critically-scales with an interaction-dependent power-law up to a finite length cutoff in the Ohmic case, (ii) get washed out in the sub-Ohmic case due to the dominant role of slow (viscous) fluctuations in the environment, and (iii) is effectively amplified as the fast charge-fluctuations (super-Ohmic) at the boundary of the wire acts similarly to a Zeno effect~\cite{Leggett87}. 
Our results highlight bath-engineering as a tool to design novel phases of matter, without the drawback of inducing dissipation.
Furthermore, we pave the way toward analyzing, e.g., the interplay between non-Ohmic reservoirs and a macroscopic number of in-wire impurities,
where another type of metal-to-insulator quantum phase transition is predicted to occur~\cite{Giamarchi04}; or the investigation of bath-induced stabilization protocols for exotic excitations, such as Majorana fermions.    

\begin{acknowledgments}
A.~K., A.~S.~, and O.~Z.~  acknowledge financial support from the Swiss National Science Foundation. A.~C.  acknowledges support by the funding from the European Research Council  (ERC) under the Horizon 2020 research and innovation program, Grant  Agreement No. 647434 (DOQS) 
\end{acknowledgments}
%

\newpage
\cleardoublepage
\setcounter{figure}{0}
\onecolumngrid

\begin{center}
\textbf{\normalsize Supplemental Material for}\\
\vspace{3mm}
\textbf{\large  A Luttinger Liquid coupled to Ohmic-class  environments}
\vspace{4mm}

{Andisheh Khedri$^{1}$, Antonio\ \v{S}trkalj$^{1}$, Alessio Chiocchetta$^{2}$, and Oded Zilberberg$^{1}$}\\
\vspace{1mm}
\textit{\small $^{1}$ Institute for Theoretical Physics, ETH Z\"urich, 8093 Z\"urich, Switzerland}\\
\vspace{1mm}
\textit{\small $^{2}$ Institute for Theoretical Physics, University of Cologne, D-50937 Cologne, Germany}
\vspace{5mm}
\maketitle

\end{center}
\setcounter{equation}{0}
\setcounter{section}{0}
\setcounter{figure}{0}
\setcounter{table}{0}
\setcounter{page}{1}
\makeatletter
\renewcommand{\bibnumfmt}[1]{[#1]}
\renewcommand{\citenumfont}[1]{#1}

\setcounter{enumi}{0}
\renewcommand{\theequation}{\Roman{enumi}.\arabic{equation}}

\section{Bosonic Green's function of a clean wire}
\setcounter{enumi}{1} 
\setcounter{equation}{0}
In this section, we outline the calculation details of the bosonic Green's function for the interacting wire. We elaborate on its crucial dependence on the environment through the boundary conditions. As mentioned in the main text, we restrict the bosonization treatment to the
interacting part of the system (wire), i.e., $x\in[-L/2,L/2]$,
and account for the presence of the environment (leads) through the following boundary conditions (continuity equation):
\begin{equation}
\partial_t \phi_{\rm L/R}(x=\pm L/2,t)=2\pi J_{\rm L/R} (t)\,,
\label{eq:boundary_condition}
\end{equation}
where the boundary operators $J_{\rm L/R}$ are the current operator in the leads
\begin{equation}
J_{\rm L/R}(\omega)=\int d\omega^\prime c^{\dagger}_{\rm L/R,\omega+\omega^\prime}c_{\rm L/R,\omega^\prime}\,,
\end{equation}
with $c^{(\dagger)}_{\rm L/R,\omega}$ the fermionic (creation) annihilation operators for right/left-moving electrons.

We use the boundary condition~\eqref{eq:boundary_condition} to solve the equations of motions
\begin{align}
&\partial_t \theta (x,t)=\frac{\nu_{\rm F}}{K_{\rm w}^2}\partial_x \varphi(x,t)\,, \label{eq:eq_of_motion_1}\\
&\partial_t \varphi(x,t)=\nu_{\rm F}\partial_x \theta (x,t)\,, 
\label{eq:eq_of_motion_2}
\end{align}
 for the bosonic fields $\varphi(x,t),\theta (x,t)=(1/\sqrt{2})\left[\phi^{\pdag}_{\rm L}(x,t)\pm \phi^{\pdag}_{\rm R}(x,t)\right]$.
We obtain a solution in the form of the original right- (left-)mover fields that reads~
\cite{sNazarov97, sAntonio19}
\begin{align}
\phi_{\rm L,R}(x,\omega)=\frac{2\pi}{i\omega}
&\Bigg\{J_{\rm L/R}(\omega)\left[\frac{4}{K_{\rm w}}\cos[\omega\tau_L(x/L\pm 1/2)
\pm 2i \left(1+\frac{1}{K_{\rm w}^2}\right)\sin[\omega\tau_{\rm L}(x/L\pm 1/2)]\right]\nonumber\\
&\mp J_{\rm R/L}(\omega)
\left[2i\left(1-\frac{1}{K_{\rm w}^2}\right)\sin[\omega\tau_{\rm L}(x/L\mp 1/2)]\right] \Bigg\}
\frac{1}{\frac{4}{K_{\rm w}}\cos[\omega\tau_L]+2i\left(1+\frac{1}{K_{\rm w}^2}\right)\sin[\omega\tau_{\rm L}]}\,,
\label{eq:fields}
\end{align}
with $\tau_{L}=L K_{\rm w}/\nu_{\rm F}$ the time-of-flight required for the bosonic excitations to cross the wire. Note that from the particle-hole symmetry of the currents at the boundaries  $J_{\rm L/R}(-\omega)=J^{\dagger}_{\rm L/R}(\omega)$, 
the following symmetry for the fields holds $\phi^{\dagger}_{\rm L/R}(x,\omega)=\phi_{\rm L/R}(x,-\omega)$. 

Using the current-current correlations at the boundaries, we can define the environment noise spectrum $S(\omega)$ as
\begin{equation}
\langle J_{\eta}(\omega)J_{\eta^{\prime}}(\omega^\prime)\rangle
=S(\omega)\delta_{\eta,\eta^\prime}\delta(\omega+\omega^\prime)\,,\hspace{1cm} \eta=\rm L,R \,.
\end{equation}
Thus, we can fully determine the correlation functions of the bosonic fields ($\varphi,\theta$) as
\begin{equation}
\mathcal{G}^{>,0}_{\varphi\varphi}(x,x^\prime,\omega)=
-i\langle \varphi(x,\omega)\varphi^{\dagger}(x^\prime,\omega^\prime) \rangle 
=-i\frac{S(\omega)\delta(\omega-\omega^\prime)}{\omega^2} F_{\varphi}(x,x^\prime,\omega)\,,
\label{eq:correlation}
\end{equation}
\begin{equation}
\mathcal{G}^{>,0}_{\theta\theta}(x,x^\prime,\omega)=
-i\langle \theta(x,\omega)\theta^{\dagger}(x^\prime,\omega^\prime) \rangle 
=-i\frac{S(\omega)\delta(\omega-\omega^\prime)}{K_{\rm w}^2\omega^2} F_{\theta}(x,x^\prime,\omega)\,,
\label{eq:correlation_theta}
\end{equation}
with the structure function
\begin{equation}
F_{\varphi/\theta}(x,x^\prime,\omega)=
\frac{\pm\left(\frac{1}{K_{\rm w}^2}-1\right)\sum_{\alpha=\pm}\cos[\omega\tau_L(x+x^\prime+\alpha L)/L]
+2\left(\frac{1}{K_{\rm w}^2}+1\right)\cos[\omega\tau_L(x-x^\prime)/L]}{\left(1+\frac{1}{K_{\rm w}^2}\right)^2-\left(1-\frac{1}{K_{\rm w}^2}\right)^2 \cos^2[\omega\tau_L]}\,,
\label{eq:structure_pt}
\end{equation}
which encodes all the information about the interacting wire, i.e., its length $L$, and the interaction strength $K_{\rm w}$. Note that, due to the presence of the environment and the correspondingly-imposed boundary, a Fabry-P\'{e}rot cavity is formed for the plasmonic excitations, with its resonances encoded in the poles of Eq.~\eqref{eq:structure_pt}. 
Setting $x^\prime=0$, the structure function takes the form
\begin{equation}
F_{\varphi/\theta}(x,0,\omega)=
\frac{K_{\rm w}^2\cos[\omega\tau_{L}x/L]}{(1+K_{\rm}^2)-(1-K_{\rm w}^2)\cos[\omega\tau_L]}\,.
\label{eq:local_struc}
\end{equation}
\subsection{Analytic properties of the structure function}

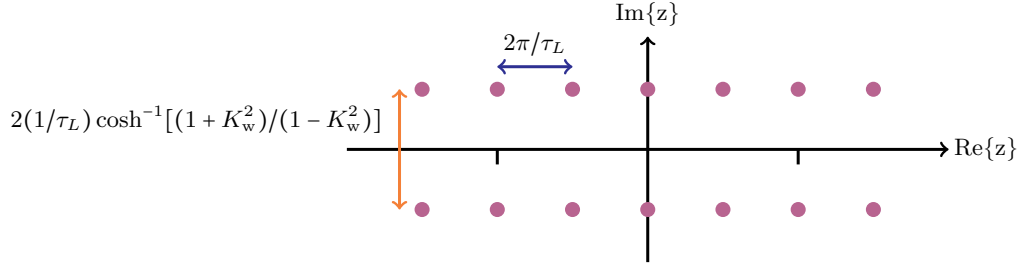
\begin{figure}[!t]
 \centering
\begin{tikzpicture}
\draw[thick,->,line width=.4mm](-4,-3)--(4,-3); 
\draw[thick,->,line width=.4mm](0,-4.5)--(0,-1.5); 
\draw[thick,-,line width=.4mm](-2,-3)--(-2,-3.2); 
\draw[thick,-,line width=.4mm](2,-3)--(2,-3.2); 
\draw(4.5,-3.0)node{$\rm{Re}\{ z\}$}; 
\draw(0.0,-1.2)node{$\rm{Im}\{ z\}$}; 
\fill[RedViolet!70!white] (0,-2.2) ellipse (0.1cm and 0.1cm);
\fill[RedViolet!70!white] (1,-2.2) ellipse (0.1cm and 0.1cm);
\fill[RedViolet!70!white] (2,-2.2) ellipse (0.1cm and 0.1cm);
\fill[RedViolet!70!white] (3,-2.2) ellipse (0.1cm and 0.1cm);
\fill[RedViolet!70!white] (-1,-2.2) ellipse (0.1cm and 0.1cm);
\fill[RedViolet!70!white] (-2,-2.2) ellipse (0.1cm and 0.1cm);
\fill[RedViolet!70!white] (-3,-2.2) ellipse (0.1cm and 0.1cm);
\fill[RedViolet!70!white] (0,-3.8) ellipse (0.1cm and 0.1cm);
\fill[RedViolet!70!white] (1,-3.8) ellipse (0.1cm and 0.1cm);
\fill[RedViolet!70!white] (2,-3.8) ellipse (0.1cm and 0.1cm);
\fill[RedViolet!70!white] (3,-3.8) ellipse (0.1cm and 0.1cm);
\fill[RedViolet!70!white] (-1,-3.8) ellipse (0.1cm and 0.1cm);
\fill[RedViolet!70!white] (-2,-3.8) ellipse (0.1cm and 0.1cm);
\fill[RedViolet!70!white] (-3,-3.8) ellipse (0.1cm and 0.1cm);
\draw[Blue,thick,<->,line width=.4mm](-2,-1.9)--(-1,-1.9);
\draw(-1.5,-1.6)node{$2\pi/\tau_L$}; 
\draw[Orange,thick,<->,line width=.4mm](-3.3,-3.8)--(-3.3,-2.2); 
\draw(-6.0,-2.6)node{$2(1/\tau_L)\cosh^{-1}[(1+K_{\rm w}^2)/(1-K_{\rm w}^2)]$}; 
\end{tikzpicture}
\caption{Schematic structure of the poles of the local structure function $F_{\varphi}(z)$, cf.~Eq.~\eqref{eq:poles}.
\label{fig:model_sketsch}}
\end{figure}

It is insightful to examine the analytic properties of the local structure function~\eqref{eq:local_struc}. For this purpose, we generalise the structure function for any complex $z\in \mathbb{C}$, and find its poles of $F_{\rm phi}$ as
\begin{equation}
z^{\pm}_{n}= \frac{1}{\tau_L}
\begin{cases}
(2n+1)\pi \pm i \cosh^{-1}\left[\frac{1+K_{\rm w}^2}{1-K_{\rm w}^2}\right]\hspace{0.5cm},\hspace{0.5cm} K_{\rm w}^2>1 \\\\
(2n)\pi \pm i \cosh^{-1}\left[\frac{1+K_{\rm w}^2}{-1+K_{\rm w}^2}\right]\hspace{0.5cm},\hspace{0.5cm} K_{\rm w}^2<1
\end{cases},
\label{eq:poles}
\end{equation}
with $n\in \mathbb{N}$, and for $F_{\theta}$ the poles are identical, but the two cases ($K_{\rm w}\lessgtr 1$) are interchanged. For $K_{\rm w}\in(0,1)$, the poles of $F_{\varphi}$ are shown in Fig.~\ref{fig:model_sketsch}.
Note that the imaginary part of the poles grows when the repulsive interaction is decreased, and as we approach the non-interacting limit $K_{\rm w}\to 1$, it becomes infinitely large.

For the purpose of performing complex integration over the structure factor~\eqref{eq:local_struc}, it is useful to rewrite it as a sum over two function
\begin{equation}
F_{\varphi}(x,0,z)
=K_{\rm w}\cos[\omega\tau_L x/L] \left\{\frac{-1}{\frac{1-K_{\rm w}}{1+K_{\rm w}}\exp[ iz \tau_L]-1}+\frac{1}{\frac{1+K_{\rm w}}{1-K_{\rm w}}\exp[ i z \tau_L]-1}
\right\}\,,
\end{equation}
where the first/second term in the parenthesis is analytic in the upper/lower-half of the complex plane, i.e., its poles are in the lower/upper-half. 
Furthermore, defining $y=\cosh^{-1}[(1+K_{\rm w}^2)/(1-K_{\rm w}^2)]$, we have $(1\pm K_{\rm w})/(1\mp K_{\rm w})=\exp(\pm y)$, and therefore we can rewrite the structure function in terms of the bosonic distribution function
\begin{equation}
F_{\varphi}(x,0,z)=K_{\rm w}\cos[\omega\tau_L x/L]\left\{-n_b\left(i\tau_L z-y\right)+n_b\left(i\tau_L z+y\right)\right\}\,.
\label{eq:decompose_F}
\end{equation}
\section{Ohmic-class environment}
\setcounter{enumi}{2} 
\setcounter{equation}{0}
We consider a generic Ohmic-class environment with noise power spectrum $S(\omega)$ defined in Eq.~(2) in the main text.
In the following, we will use the retarded bosonic Green's function for the field $\varphi$ of the clean wire, 
$\mathcal{G}^{\rm R,0}_{\varphi \varphi}(x,t;x^\prime,t^\prime)=
-i\Theta(t-t^\prime)\langle [\varphi(x,t),\varphi^\dagger(x^\prime,t^\prime)]\rangle, $
with $\Theta$ the Heaviside function. It can be expressed in terms of lesser and greater Green's function
$\mathcal{G}^{<,0}_{\varphi \varphi}(x,t;x^\prime,t^\prime)=
-i\left\langle \varphi^{\dagger}(x^\prime,t^\prime) \varphi(x,t)\right\rangle,
$
and 
$\mathcal{G}^{>,0}_{\varphi \varphi}(x,t;x^\prime,t^\prime)=
-i\left\langle \varphi(x,t)\varphi^{\dagger}(x^\prime,t^\prime)\right\rangle,
$
as
\begin{equation}
\mathcal{G}^{\rm R,0}_{\varphi \varphi}(x,x^\prime,\omega)=
i\int_{-\infty}^{\infty}\frac{d\omega^\prime}{2\pi}
\frac{\mathcal{G}^{>,0}_{\varphi\varphi}(x,x^\prime,\omega^\prime)-\mathcal{G}^{<,0}_{\varphi\varphi}(x,x^\prime,\omega^\prime)}
{\omega^\prime-\omega-i 0^{+}}\,.
\label{eq:retarded_D}
\end{equation}
Note that Ohmic-class environments hold detailed balance, and we have
\begin{equation}
\mathcal{G}^{<,0}_{\varphi\varphi}(x,x^\prime,\omega)=
\frac{S(-\omega)}{S(\omega)}\mathcal{G}^{>,0}_{\varphi\varphi}(x,x^\prime,\omega)
=
e^{-\beta\omega}\mathcal{G}^{>,0}_{\varphi \varphi^\prime}(x,x^\prime,\omega)\,,
\label{eq:DG_DL_relation}
\end{equation}
which follows from Eq.~\eqref{eq:correlation} [we use the fact that $F_{\varphi}(x,x^\prime,\omega)$ is a real and even function of $\omega$, see Eq.~\eqref{eq:structure_pt}], as we expect for bosons in thermal equilibrium.  
Thereby, the retarded Green's function Eq.~\eqref{eq:retarded_D} simplifies to
\begin{equation}
\mathcal{G}^{\rm R,0}_{\varphi\varphi}(x,x^\prime,\omega)=
\int_{-\infty}^{\infty}\frac{d\omega^\prime}{2\pi}\frac{F_{\varphi}(x,x^\prime,\omega^\prime)}{\omega^\prime-\omega-i 0^{+}}~
\left|\frac{\omega^\prime}{\omega_c}\right|^{s-1}~ \frac{e^{-|\omega^\prime/\omega_c|}}{\omega^\prime}\,.
\label{eq:detailed_retarded}
\end{equation}

\subsection{Ohmic-environment}
In the case of an Ohmic environment, $s=1$, we extend the integral in Eq.~\eqref{eq:detailed_retarded} to the complex plane, and employ the analytic properties of the structure function [see discussion leading to Eq.~\eqref{eq:decompose_F}], to obtain the local retarded Green's function for $x^\prime=0$
\begin{equation}
\mathcal{G}_{\varphi\varphi}^{\rm R/A,0}(x,0,\omega)=
\frac{\pm iK_{\rm w}/2}{\omega\pm i\eta}
\left\{
\frac{e^{-i\tau_L(\omega\pm i\eta)x/L}+\frac{1-K_{\rm w}}{1+K_{\rm w}}e^{i\tau_L(\omega\pm i\eta)(x-L)/L}}{1-\frac{1-K_{\rm w}}{1+K_{\rm w}}e^{i\tau_L(\omega\pm i\eta)}}
\right\}\,.
\label{eq:retarded_D_trc}
\end{equation}
Analytically continuing the obtained retarded/advanced Green's function to the imaginary axis (Matsubara space) \cite{sbruus04} by substitution $\omega\pm i\eta \to i\omega$ for $\omega\gtrless 0$, we obtain the Matsubara Green's function.
The the same results can be obtained by solving the equation of motion for Matsubara Green's function 
\begin{equation}
\frac{1}{k(x)}\left[-\frac{\nu_{\rm F}}{k(x)}\partial_x^2+\frac{k(x)\omega^2}{\nu_{\rm F}}\right] 
\mathcal{G}^{0}_{\varphi\varphi}(x,x^\prime,i\omega)=\delta(x-x^\prime),
\end{equation}
where $k(x)=K_{\rm w}$ for $x\in[-L/2,L/2]$, and $k(x)=1$ elsewhere.
This is accomplished by imposing the continuity of the Matsubara Green's function and of its derivative $v_{\rm F}/k^2(x)\partial_x \mathcal{G}_{\varphi\varphi}^{0}(x,x^\prime,i\omega)$ at the wire's ends $x=\pm L/2$, as well as at the discontinuity at $x=x^\prime$, namely 
\begin{equation}
\frac{\nu_{\rm F}}{k^2(x)}\partial_x \mathcal{G}^{0}_{\varphi\varphi}(x,x^\prime,i\omega)|_{x=x^\prime-0^+}^{x=x^\prime+0^+}=-1\,.
\end{equation}

Note that in the limit of $L\to\infty$, i.e., far from the boundaries, the Matsubara Green's function reads $\mathcal{G}^0_{\varphi\varphi}(x,x^\prime,i\omega)\approx\mathcal{G}^0_{\varphi\varphi}(x-x^\prime,0,i\omega)=K_{\rm w}/(2|\omega|) \exp[-\nu|(x-x^\prime)\omega|]$, with $\nu=\nu_{\rm F}/K_{\rm w}$ resulting in
\begin{equation}
\mathcal{G}^0(q,i\omega)=\int dx e^{iqx}\mathcal{G}^0(x,0,i\omega)=\frac{K_{\rm w}}{2|\omega|}\frac{1}{\nu q^2+\omega^2/\nu}.    
\end{equation}

For a finite $L$, the analytical continuation of Eq.~\ref{eq:retarded_D_trc} reads $\mathcal{G}^{0}_{\varphi\varphi}(i\omega)=K(\omega)/(2|\omega|)$, with
\begin{equation}
K(\omega)=K_{\rm w}\frac{1+\frac{1-K_{\rm w}}{1+K_{\rm w}}e^{-\tau_L|\omega|}}{1-\frac{1-K_{\rm w}}{1+K_{\rm w}}
e^{-\tau_{L}|\omega|}}\,.
\label{eq:kLL_omega}
\end{equation}
At large frequencies, $\omega\tau_L\gg 1$, $K(\omega)$ approaches the Luttinger Liquid parameter inside the wire $K_{\rm w}$, whereas for low frequencies, $\omega\tau_L\ll 1$, $K(\omega)$ goes to $1$, resembling the TLL parameter of the noninteracting (Fermi liquid) leads.
In conclusion, the finite length of the wire connected to non-interacting electrons introduces an infrared cutoff $1/\tau_L$, below which the wire acts as non-interacting electrons.
\subsection{Infinite wire connected to an Ohmic-class environment}
\begin{figure}[!htbp]
 \centering
\includegraphics[width=1.0\linewidth]{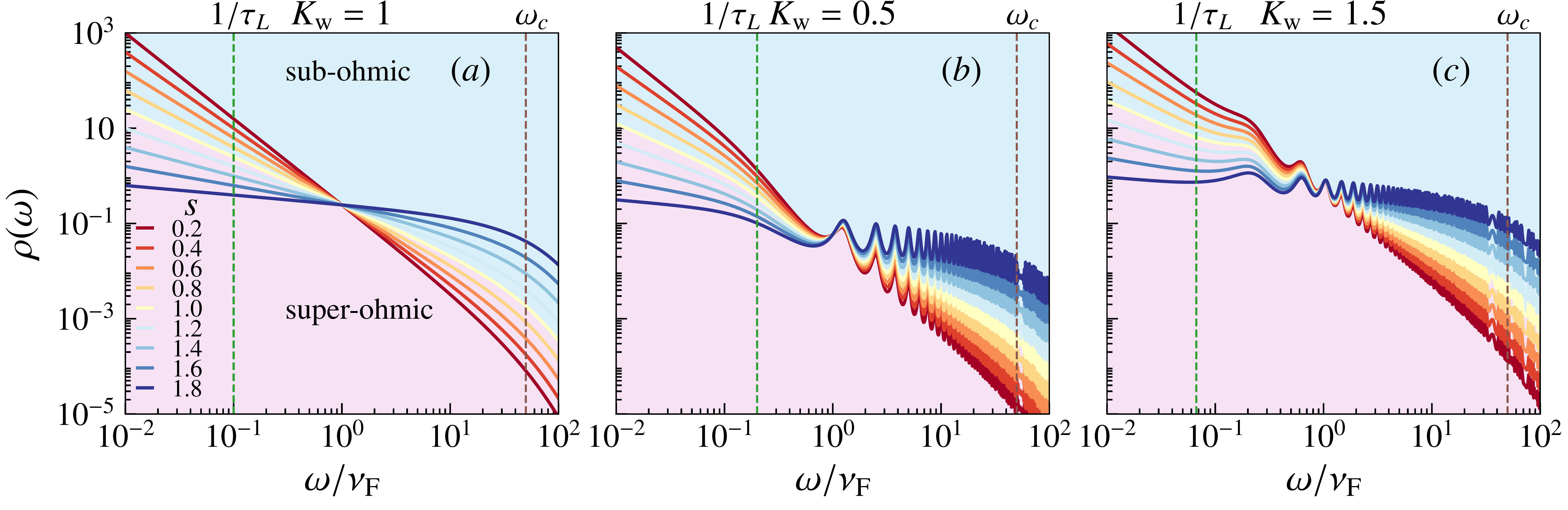}
 \caption{The bosonic spectral function at $x=x^\prime=0$ as a function of frequency for different values of Ohmic-class noise $s$ and interaction strength $K_{\rm w}$ for a finite wire of length $L/\nu_{\rm F}=10$. Vertical dashed lines mark frequencies $1/\tau_{\rm L}$ (green) below which the finite length of the wire plays a dominant role, and $\omega_c$ (red) above which the environment noise is cut off. }
 \label{fig:plasmonic_spec}
\end{figure}

We next consider a generic Ohmic-class environment with $s\in(0,2)$, for which the bosonic spectral function reads
\begin{equation}
\rho(x,x^\prime,\omega)=\text{Im}\{\mathcal{G}^{\rm R,0}_{\varphi\varphi}(x,x^\prime,\omega)\}= F_{\varphi}(x,x^\prime,\omega) 
\frac{\text{sgn}(\omega)}{2\omega_c} \left|\frac{\omega}{\omega_c}\right|^{s-2}~ e^{-|\omega/\omega_c|},
\end{equation}
as is shown in Fig.~\ref{fig:plasmonic_spec}. In the super-Ohmic case, the spectral function at small frequencies, $\omega\tau_{\rm L}\ll 1$, diverges slower than the ohmic case [i.e., slower than $\sim K_{\rm w}/(2\omega)$], while the sub-Ohmic case is diverging faster. For the interacting wire at frequencies $\omega\tau_{\rm L}>1$, the oscillations are present due to the formation of a Fabry-P\'{e}rot cavity. 

Having the plasmonic spectral function, we can obtain the plasmonic Matsubara Green's function, at Matsubara frequencies $\omega_n=2n \pi/\beta$, $n\in \mathbb{N}$,
\begin{align}
\mathcal{G}^0_{\varphi\varphi}(x,0,i\omega_n)&=\int_{-\infty}^{\infty}\frac{d\omega^\prime}{\pi} ~   
\frac{\rho(\omega^\prime )}{\omega^\prime-i\omega_n} 
=\frac{-K_{\rm w}\omega_c^{1-s}}{2\sin[s\pi/2]}\text{Im}\left\{\sum_{\alpha,\alpha^\prime=\pm}
\int_{\mathcal{C}_{\alpha^\prime}} \frac{dz}{2\pi} \frac{(-i z)^{s-2}}{z-i\omega_n} \frac{\alpha e^{\alpha i\tau_L z x/L}}{\frac{1-\alpha K_{\rm w}}{1+\alpha^\prime\alpha K_{\rm w}}e^{\alpha i\tau_L z}-1}
\right\},\nonumber\\
&=\frac{-K_{\rm w}}{2\omega_c\sin[s\pi/2]}\left|\frac{\omega_n}{\omega_c}\right|^{s-2}
\frac{e^{-\tau_L|\omega_n x/L|}+\frac{1-K_{\rm w}}{1+K_{\rm w}}e^{\tau_L|\omega_n|(x-L)/L}}{1-\frac{1-K_{\rm w}}{1+K_{\rm w}}e^{-|\omega_n|\tau_L}}\,,
\label{eq:G_L_inf}
\end{align}
with $\mathcal{C}_{\pm}$ being a half-circle contour extending the real axis to the upper/lower half of the complex plane. We used the fact that we are interested in $\omega_n\ll\omega_c$.
Taking the limit of $L\to\infty$, we can rewrite Eq.~\eqref{eq:G_L_inf}, as
$\mathcal{G}^{0}_{\varphi\varphi}(x,0,i\omega)=K(\omega)/(2|\omega|)e^{-|\omega x/\nu|}$, with $K(\omega)=K_{\rm w}\text{Csc}[(\pi s)/2]|\omega/\omega_c|^{s-1}$.
In the Ohmic case, this reduces to $K(\omega)=K_{\rm w}$ as expected.
Analogous to the Ohmic case, we obtain $\mathcal{G}_{\varphi\varphi}(q,i\omega_n)=K(\omega)/[\nu q^2+\omega_n^2/\nu]$.

\subsection{Finite wire connected to an Ohmic-class environment}
\begin{figure}[!htbp]
 \centering
\includegraphics[width=1.0\linewidth]{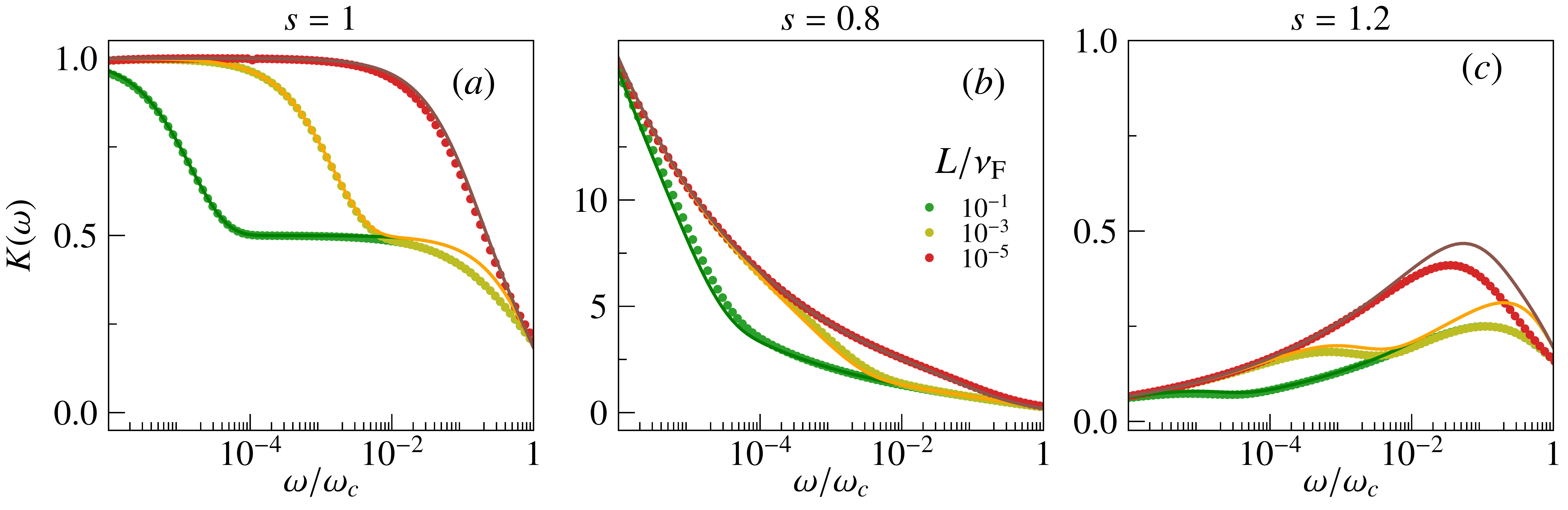}
 \caption{The frequency dependence of effective TLL parameter $K(\omega)$ for various lengths of the interacting wire with $K_{\rm w}=0.5$. Filled circles represent the numerical calculation of Eq.~\eqref{eq:kLL_omega_num}, and the solid lines show the approximate formula~\eqref{eq:kLL_omega_2}. }
 \label{fig:kLL_omega}
\end{figure}

In the case of a finite wire, similar to the infinite length case, we can define the effective TLL parameter as
\begin{equation}
K(\omega)= \omega\int_{0}^{\infty}\frac{d\omega^\prime}{\pi} ~   
\frac{F_{\varphi}(0,0,\omega^\prime)}{\omega^2+{\omega^\prime}^2} 
\frac{\text{sgn}(\omega^\prime)}{\omega_c} \left|\frac{\omega^\prime}{\omega_c}\right|^{s-2}~ e^{-|\omega^\prime/\omega_c|}\, .
\label{eq:kLL_omega_num}
\end{equation} 
In Fig.~\ref{fig:kLL_omega}, we show the numerical evaluation of the integral above, and compare it with the approximate formula
\begin{equation}
K(\omega)=\frac{K_{\rm w}}{\sin[(\pi s)/2]}\left|\frac{\omega}{\omega_c}\right|^{s-1}\frac{1+K_{\rm w}+(1-K_{\rm w})e^{-\tau_{\rm L}|\omega|}}{1+K_{\rm w}-(1-K_{\rm w})e^{-\tau_{\rm L}|\omega|}}\,.
\label{eq:kLL_omega_2}
\end{equation}
The approximation agrees well for $\omega\ll\omega_c$, irrespective of the value of $s$.
Note that from Eq.~\eqref{eq:kLL_omega_2} it follows that at high frequencies, $\omega\tau_{\rm L}\gg 1$, $K(\omega)=K_{\rm w}|\omega/\omega_c|^{s-1}/\sin(s\pi/2)$. 
At small frequencies, $\omega\tau_{\rm L} \ll 1$, on the other hand, 
the curves corresponding to different lengths merge together as is shown in Fig.~\ref{fig:kLL_omega}, exhibiting that at such small frequencies, the physics of the interacting wire is washed out and the system's behaviour is dominated by the environment. 

\section{Renormalization of the scattering potential}
\setcounter{enumi}{3} 
\setcounter{equation}{0}
In this section, we present the standard perturbative RG analysis for the problem of a single impurity immersed in a TLL. After integrating-out the leads, the action of the system in imaginary-time reads
\begin{align}
\resizebox{.91\hsize}{!}{$\displaystyle{\mathcal{A}_{\rm s}[\{\varphi\}]
=\frac{1}{4\pi}
\int_{0}^{\beta}d\tau \int_{0}^{\beta}d\tau^\prime
\int_{-L/2}^{L/2}{dx}\int_{-L/2}^{L/2}{dx'} 
\varphi^{\dagger}(x,\tau) \mathcal{G}^{0,-1}_{\varphi\varphi}(x,\tau;x^\prime,\tau^\prime) \varphi(x^\prime,\tau')
+V_0\int_{0}^{\beta} d\tau~\cos\left[\gamma \varphi(x_0,\tau)\right]\,,}$}
\label{eq:action_system}
\end{align}
with $\gamma=\sqrt{4\pi}$, and $\varphi(\tau)=\int_{-\infty}^{\infty}~ d\omega\, e^{i\omega\tau} \varphi(\omega)$. For all $x\neq x_0$, the action is quadratic, and we can integrate out all the corresponding fields to obtain the local action [Eq.~(4) in the main text], which can be decomposed as
$\mathcal{A}[\varphi]=\mathcal{A}_0[\varphi]+\mathcal{A}_{\rm imp}[\varphi]$,
with  
\begin{align}
\mathcal{A}_0[\phi]&=\int_{0}^{\beta}d\tau~\varphi^\dagger (\tau)
\mathcal{G}^{0,-1}_{\varphi\varphi}(\tau) \varphi^{\pdag} (\tau), \nonumber\\
\mathcal{A}_{\rm imp}[\phi]&=
V_0\int_{0}^{\beta}d\tau \cos[\gamma\varphi(\tau)],
\label{eq:action_local}
\end{align}
where we have assumed $x_0=0$.
We define an ultraviolet cut-off $\Lambda$ and the corresponding scale-dependent field as $\varphi_{\Lambda}(\tau)=\int_{-\Lambda}^{\Lambda}~ d\omega \,e^{i\omega\tau} \varphi(\omega)$. For any $\Lambda^\prime\in[0,\Lambda]$, we can decompose the bosonic fields into low- and high-frequency fields, $\varphi_{\Lambda}(\tau)=\varphi_{\Lambda^\prime}(\tau)+h(\tau)$.
Now, we turn to the functional-integral formulation of the partition function, and integrate over the high-frequency field $h(\tau)$ to obtain $\mathcal{Z}=\int \mathcal{D}[\{\varphi_{\Lambda^\prime}\}] e^{-\mathcal{A}_{\rm eff}[\{\varphi_{\Lambda^\prime}\}]},$
with
\begin{align}
\mathcal{A}_{\rm eff}[\{\varphi_{\Lambda^\prime}(\tau)\}]
&=\mathcal{A}_0[\{\varphi_{\Lambda^\prime}(\tau)\}]
+\left\langle \mathcal{A}_{\rm imp}[\{\varphi_{\Lambda^\prime}(\tau)+h(\tau)\}]  \right\rangle_{\{h(\tau)\}}
+\left\langle \mathcal{A}^2_{\rm imp}[\{\varphi_{\Lambda^\prime}(\tau)+h(\tau)\}]  \right\rangle_{\{h(\tau)\}}
+\cdots,
\end{align}
where
\begin{align}
\left\langle \mathcal{A}_{\rm imp}[\{\varphi_{\Lambda^\prime}(\tau)+h(\tau)\}]  \right\rangle_{\{h(\tau)\}}
&=V_0\int_{0}^{\infty}d\tau\int \mathcal{D}[\{h(\tau)\}]e^{-\mathcal{A}_{0}[\{h(\tau)\}]}
\cos\left[\gamma\left(\varphi_{\Lambda^\prime}(\tau)+h(\tau)\right)\right]\nonumber\\
&=\frac{V_0}{2}\int_{0}^{\infty}d\tau 
\left\{ e^{i\gamma \varphi_{\Lambda^\prime}(\tau)}
\int \mathcal{D}[\{h(\tau)\}] e^{-\mathcal{A}_{0}[\{h(\tau)\}]+i\gamma h(\tau)} 
+\text{c.c}\right\}.
\end{align}
Performing the Gaussian integral over high-frequency fields we obtain
\begin{align}
\left\langle \mathcal{A}_{\rm imp}[\{\varphi_{\Lambda^\prime}(\tau)+h(\tau)\}]  \right\rangle_{\{h(\tau)\}}=
\frac{V_0}{2} e^{-\gamma^2\int_{\Lambda'}^{\Lambda} \frac{d\omega}{2\pi}~ \mathcal{G}^{0}_{\varphi\varphi}(i\omega)}
\left\{
e^{i\gamma\varphi_{\Lambda^\prime}}
+\text{c.c}
\right\} \, .
\end{align}

We proceed with the RG procedure and consider an infinitesimal change of the ultraviolet cutoff $\Lambda^\prime=\Lambda-d\Lambda=\Lambda(1-dl)$, $dl=d\Lambda/\Lambda$. In order to compare the physics governed by fields $\varphi_{\Lambda'}$ with the one governed by the fields $\varphi_{\Lambda}$, we re-scale the frequency $\omega$ to $\omega^\prime=\omega \, \Lambda/\Lambda^\prime=(1+dl)\omega$, and the corresponding imaginary-time $\tau$ to $\tau^\prime=\tau(1-dl)$, such that $\tau\omega=\tau^\prime\omega^\prime$~\cite{sgogolin2004}. Thereby, we have
\begin{align}
\left\langle \mathcal{A}_{\rm imp}[\{\varphi_{\Lambda^\prime}(\tau)+h(\tau)\}]  \right\rangle_{\{h(\tau)\}}=  
\frac{V_0}{2} e^{-\gamma^2\int_{\Lambda'}^{\Lambda} \frac{d\omega}{2\pi}~ \mathcal{G}^{0}_{\varphi\varphi}(i\omega)} (1+dl)\int~d\tau^\prime \cos[\gamma\varphi_{\Lambda^\prime}(\tau^\prime)] \, .
\end{align}
Now the high-frequency degrees of freedom can be integrated out while keeping the partition function invariant   
\begin{align}
\mathcal{Z}&=\int \mathcal{D}[\{\phi_{\Lambda}\}]
e^{-\mathcal{A}_{0}[\{\phi_{\Lambda}\}]
-\mathcal{A}_{\rm imp}[\{\phi_{\Lambda}(\tau),V_0\}]}\\
&=\int \mathcal{D}[\{\phi_{\Lambda(1-dl)}\}]
e^{-\mathcal{A}_{0}[\{\phi_{\Lambda(1-dl)}\}]
-\mathcal{A}_{\rm imp}[\{\phi_{\Lambda(1-dl)},V(\Lambda)\}]}+\mathcal{O}\left(V_0^2\right)\,,
\end{align}
provided that the scattering potential is renormalized accordingly, i.e.,
\begin{equation}
V(\Lambda)=V_0\left[ 1+e^{-\gamma^2\int_{\Lambda'}^{\Lambda}\frac{d\omega}{2\pi}\mathcal{G}^0_{\varphi\varphi}(i\omega)}\left(1+dl\right)\right] \, .
\label{eq:flow_general}
\end{equation}
Using the plasmonic Matsubara Green's function for an infinitely long-wire coupled to the Ohmic-class environment that we obtained in Eq.~\eqref{eq:G_L_inf}, the integral in the exponent becomes
\begin{align}
\int_{\Lambda'}^{\Lambda} \frac{d\omega}{2\pi}\mathcal{G}^{0}_{\varphi\varphi}(i\omega)&=   
K_{\rm w}\text{Csc}[s\pi/2]\int_{\Lambda(1-dl)}^{\Lambda}~ \frac{d\omega}{\omega_c}~ e^{-\omega/\omega_c}
\left|\frac{\omega}{\omega_c}\right|^{s-2} \nonumber\\
&=K_{\rm w}~\text{Csc}[s\pi/2]~\left\{\Gamma[s-1,\Lambda/\omega_c]-\Gamma[s-1,(\Lambda/\omega_c)(1-dl)]\right\} \nonumber\\
&=K_{\rm w}~\text{Csc}[s\pi/2]
\left(\frac{\Lambda}{\omega_c}\right)^{s-1} e^{-\Lambda/\omega_c} ~ dl
+\mathcal{O}[(dl)^2] \, ,
\end{align}
where $\Gamma$ is the incomplete Gamma function. Therefore the flow equation up to the first order in $dl$ is 
\begin{equation}
\frac{dV}{dl}=V_0\left[1-K_{\rm w}~\text{Csc}[s\pi/2]
\left(\frac{\Lambda}{\omega_c}\right)^{s-1} e^{-\Lambda/\omega_c}
\right] \, .  
\label{eq:flow_gneral}
\end{equation}
Taking the limit $\omega_c\to\infty$, in the Ohmic-case ($s=1$), the flow equation boils down to $dV/dl=V_0(1-K_{\rm w})$, which is the known result from Kane and Fisher~\cite{sKane92b}.

\begin{figure}[!htbp]
 \centering
\includegraphics[width=1.0\linewidth]{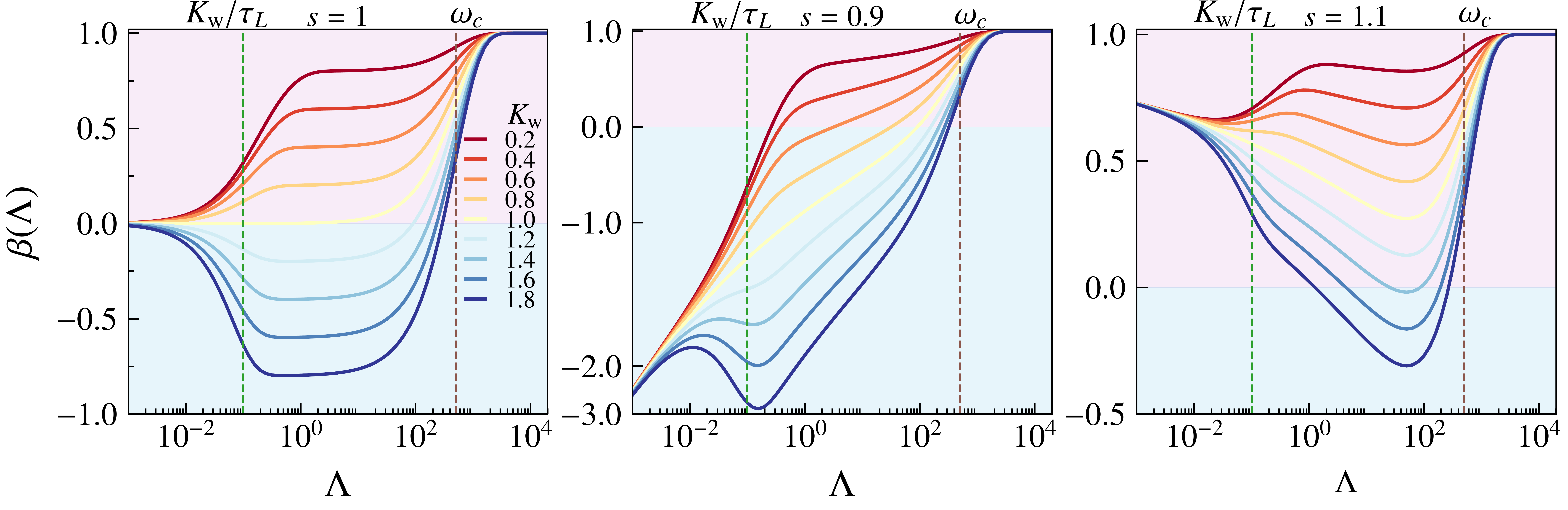}
 \caption{The beta function, i.e., the right-hand side of Eq.~\eqref{eq:flow_gneral}, as a function of frequency cut-off for different values of $s$ and interaction strength $K_{\rm w}$. Vertical dashed lines mark frequencies $K_{\rm w}/\tau_{\rm L}$ (green) below which the beta function becomes independent of the interaction strength in the wire, and $\omega_c$ (red) above which the environment noise is cut off. }
 \label{fig:beta}
\end{figure}

For a finite-length wire, we can obtain the flow equation for the scattering potential in an analogous manner. 
Figure \ref{fig:beta} depicts the beta function $dV/dl=\beta(\Lambda)$ as a function of scale parameter $\Lambda$ for the Ohmic, sub-Ohmic, and the super-Ohmic cases, and for varying interaction strength $K_{\rm w}$. As it is shown for $\Lambda\tau_L\ll 1$, the beta function is independent of the interaction strength in the wire, which is consistent to our previous discussion in relation to Eq.~\eqref{eq:kLL_omega_2}. The flow for $\Lambda\tau_L\gg 1$ is similar to our analysis of the infinitely long wire in the main text.


\section{Conductance}
\setcounter{enumi}{4} 
\setcounter{equation}{0}
In this section, we consider an external electrical potential $U(x,t)$, and calculate the resulting conductance through the interacting wire containing an impurity. The action of the system in imaginary time reads
\begin{align}
\mathcal{A}_{\rm tot}[\{\varphi\}]
=\mathcal{A}_{\rm s}[\{\varphi\}]+\int_{0}^{\beta}d\tau \int_{-L/2}^{L/2} dx~ E(x,\tau)~\varphi(x,\tau) \, ,
\label{eq:action_con}
\end{align}
where $\mathcal{A}_{\rm s}$ is given in Eq.~\eqref{eq:action_system}, and $E(x,t)=-\partial_x U(x,t)$ is the external electric field. Similar to the previous section, for all $x\neq x_0$, the action is quadratic, and we can integrate out all the fields at $x\neq x_0$~\cite{sFurusaki96}, and obtain the following local action
\begin{align}
\mathcal{A}_{\rm eff}=&\frac{1}{\beta}\sum_{\omega_n}\varphi(-\omega_n)\mathcal{G}_{\varphi\varphi}^{0,-1}(x_0,x_0,i\omega_n)\varphi(\omega_n)  
+V_0\int_{0}^{\beta} d\tau~\cos[{\gamma \varphi(\tau)}]\nonumber\\
&+\sum_{\omega_n}~ \mathcal{G}_{\varphi\varphi}^{0,-1}(x_0,x_0,i\omega_n) \varphi(-i\omega_n)\int dx~ \mathcal{G}^0_{\varphi\varphi}(x,x_0) E(x,\omega_n)\varphi(-i\omega_n)\nonumber\\
&-\frac{\beta}{4}\sum_{\omega_n}\int dx \int dx^\prime \left[\mathcal{G}^0_{\varphi\varphi}(x,x^\prime,i\omega_n)
-\mathcal{G}^0_{\varphi\varphi}(x,0,i\omega_n) \mathcal{G}^0_{\varphi\varphi}(x_0,x^\prime,i\omega_n)\left[\mathcal{G}^0_{\varphi\varphi}(0,0)\right]^{-1} \right] E(x,\omega_n)E(x^\prime,-\omega_n) \, .
\end{align}
In the limit $\omega_n \to 0$, the structure function~\eqref{eq:structure_pt} and hence the Matsubara Green's function $\mathcal{G}^0(i\omega_n)$
become independent of position ($x,x^\prime$), and hence the local action can be approximated as
\begin{align}
\mathcal{A}_{\rm eff}=&\frac{1}{\beta}\sum_{\omega_n}\varphi(-\omega_n)[\mathcal{G}_{\varphi\varphi}^{0}(i\omega_n)]^{-1}\varphi(\omega_n)    
+V_0\int_{0}^{\beta} d\tau~\cos[{\gamma \varphi(\tau)}]
+\sum_{\omega_n} \varphi(-i\omega_n)U\cos(\omega t),
\end{align}
where we have assumed $x_0=0$, and $\int_{-L/2}^{L/2} dx~ E(x,\omega_n)=U\cos(\omega t)$. 
We are interested in calculating the resulting ac-current, which reads~\cite{sGlazman1997}
\begin{align}
\frac{I_{\rm ac}}{2(e^2/h)\omega}=\langle\phi(t)\rangle=
\int \mathcal{D}[\{\varphi\}]~ \phi(\tau)~ e^{-\mathcal{A}_{\rm eff}}=
U\int dt^\prime ~ \mathcal{G}_{\varphi\varphi}(t,t^\prime) \cos(\omega t^\prime) 
=U~\text{Re}\left\{ e^{i\omega t} \mathcal{G}_{\varphi\varphi}(i \omega)\right\},
\label{eq:ac_current}
\end{align}
with $\mathcal{G}_{\varphi\varphi}$ being the plasmonic Matsubara Green's function in the presence of the impurity. 
As the exact form of the $\mathcal{G}_{\varphi\varphi}$ is unattainable, we perform a perturbative expansion~\cite{sKane92b} in terms of the scattering potential $V_0$, and obtain 
\begin{equation}
\mathcal{G}_{\varphi\varphi}(i\omega_n)=
\int \mathcal{D}[\{\varphi\}]~ \phi(-\omega_n)\phi(\omega_n)~ e^{-\mathcal{A}_{\rm eff}}
=
\mathcal{G}^0_{\varphi\varphi}(i\omega_n)+\mathcal{G}^0_{\varphi\varphi}(i\omega_n) \Sigma(i\omega_n)\mathcal{G}^0_{\varphi\varphi}(i\omega_n)
+\mathcal{O}(V_0^4) \, ,
\end{equation}
where
\begin{equation}
\Sigma(i\omega_n)=
-V_0^2\frac{\gamma^2}{2}
\sum_{\alpha_1,\alpha_2=\pm}\int_{0}^{\beta} d\tau_1 \int_{0}^{\beta} d\tau_2
\left[1+\alpha_1 \alpha_2 \cos\left[\omega_n(\tau_1-\tau_2)\right]\right]
e^{-\frac{\gamma^2}{2\beta}\sum_{\omega_n} \mathcal{G}_{\varphi\varphi}^{0}(i\omega_n)[1+\alpha_1\alpha_2\cos[\omega_n(\tau_1-\tau_2)]]} 
\label{eq:I_sum}
\end{equation}
is the impurity-induced self energy. 
In the following, we try to simplify the above expression by using the analytic properties of Matsubara Green's functions.
First, we define
\begin{equation}
E(\tau)=-\frac{2\gamma^2}{\beta}\sum_{\omega_n} ~\mathcal{G}_{\varphi\varphi}^{0}(i\omega_n) [1-\cos(\omega_n\tau)]
=-\frac{\gamma^2}{\beta}\sum_{\omega_n>0}~\mathcal{G}_{\varphi\varphi}^{0}(i\omega_n) \left[1+e^{i\omega_n\beta}-e^{i\omega_n\tau}-e^{i\omega_n(\beta-\tau)}\right] \, , 
\label{eq:E_tau_sum}
\end{equation}
where we have used $e^{i\omega_n\beta}=1$.
Using the Bose-Einstein distribution function $n_b(z)=[\exp(\beta z)-1]^{-1}$, we can rewrite the sum in Eq.~\eqref{eq:E_tau_sum} as the contour integral in the complex plane. Furthermore, we deform the contour to be parallel to the real axis, and obtain
\begin{equation}
E(\tau)= -\gamma^2 \int_{-\infty}^{\infty} \frac{d\omega}{\pi i} ~\mathcal{G}_{\varphi\varphi}^{0,\rm R}(\omega)
\left\{\coth(\beta\omega/2)[1-\cosh(\omega\tau)]+\sinh(\omega\tau)\right\} \, .
\label{eq:E_tau_simp}
\end{equation}
Therefore, the impurity-induced self energy becomes
\begin{equation}
\Sigma(i\omega_n)=V_0^2\int_{0}^{\beta} d\tau~ \left[1-\cos(\omega_n\tau)\right]e^{E(\tau)}
=V_0^2\int_{0}^{\beta} d\tau \left[ 1-\exp(i\omega_n\tau)\right]e^{E(\tau)} \, ,
\end{equation}
where we have used the periodic properties of $E(\tau)=E(\beta-\tau)$.
Further deforming of the contour integration in the complex plane results in
\begin{align}
\Sigma(i\omega_n)=iV_0^2\left\{
\int_{0}^{\infty} dt ~\left[1-e^{-\omega_n t}\right] e^{E(it)}
- \int_{-\infty}^{0} dt ~\left[1-e^{\omega_n t}\right] e^{E(it)}
\right\} \, . 
\end{align}
Now, we perform the analytical continuation to the real axis ($i \omega_n\to \omega+ i 0^+$), and obtain
\begin{equation}
\Sigma^{\rm R}(\omega)=i V_0^2
\left\{\int_{0}^{\infty} dt~\left[1-e^{it(\omega+i0^+)}\right]e^{E(it)}
-i \int_{-\infty}^{0} dt~\left[1-e^{-it(\omega+i0^+)}\right]e^{E(it)} \right\}
=i \omega \tilde{I}+\mathcal{O}(\omega^2) \, ,
\label{eq:I_R}
\end{equation}
with 
\begin{equation}
\tilde{I}=-i V_0^2\int_{-\infty}^{\infty} dt~ t~ e^{E(it)}=
- V_0^2 (\beta/2) \int_{-\infty}^{\infty} dt~e^{E(it)} \, .
\label{eq:I_tilde}
\end{equation}
Note that in Eq.~\eqref{eq:I_tilde}, we once again employed the periodic properties $E(\beta-it)=E(it)$.
Finally, the retarded Green's function up to second order in $V_0$ reads
\begin{equation}
\mathcal{G}^{\rm R}_{\varphi\varphi}(\omega)=
\mathcal{G}^{\rm R,0}_{\varphi\varphi}(\omega)
[1+ i \omega \tilde{I} \mathcal{G}^{\rm R,0}_{\varphi\varphi}(\omega)] \, ,
\end{equation}
which can be employed to calculate the ac current Eq.~\eqref{eq:ac_current}, and hence the ac conductance
\begin{align}
G(\omega)=\frac{I_{\rm ac}(\omega)}{U\cos(\omega t)} 
=2\frac{e^2}{h} \left\{ i \omega \mathcal{G}^{\rm R,0}_{\varphi\varphi}(\omega)
- \omega^2 \tilde{I} \left[\mathcal{G}^{\rm R,0}(i\omega) \right]^2
\right\} \, .
 \label{eq:conductance_simp}
\end{align}
The first term in Eq.~\eqref{eq:conductance_simp} is the conductance of a clean wire [$G_0$ in the main text], and the second term corresponds to the back-scatterings from the impurity [$G_b$ in the main text].
We conclude this section by emphasising that the calculation of the conductance~\eqref{eq:conductance_simp} boils down to the evaluation of the two integrals in Eqs.~\eqref{eq:I_tilde} and~\eqref{eq:E_tau_simp}.
\subsection{Ohmic environment}
In this subsection, we outline the analysis of the temperature dependence of the conductance through the wire which is in contact with Ohmic leads with noise power spectrum $S(\omega)=\omega \left[1+n_b(\beta\omega)\right]$.

\subsubsection{Infinitely long wire}
For an infinitely long-wire, the local retarded Green's function is $\mathcal{G}^{0,\rm R}_{\varphi\varphi}(i\omega)=\frac{iK_{\rm w}}{2\omega+i0^+}e^{-|\omega/\omega_c|}$. Hence, Eq.~\eqref{eq:E_tau_simp} becomes
\begin{align}
E(i t)&=-\frac{K_{\rm w}\gamma^2}{\pi} \int_{0}^{\infty} d\omega~ \frac{e^{-\omega/\omega_c}}{\omega} \left\{ [1-\cos(\omega t)]\coth(\beta\omega)+i\sin(\omega t)\right\} \nonumber \\
&=-\frac{K_{\rm w}\gamma^2}{\pi} \ln[1+i\omega_c t]
+\sum_{m=1}^{\infty}\ln\left[1+\left(\frac{t}{m\beta+(1/\omega_c)}\right)^2\right] \, .
\end{align}
Concentrating on the low temperature behaviour, $\omega_c\beta\gg 1$, we obtain
\begin{equation}
E(it)\approx -\frac{K_{\rm w}\gamma^2}{\pi}\ln\left\{
\left[1+i\omega_c t\right
]\left[\frac{\beta}{\pi t}\sinh\left( \frac{\pi t}{\beta}\right)\right]
\right\} \, ,   
\end{equation}
which makes the evaluation of Eq.~\eqref{eq:I_tilde} feasible, resulting in
\begin{equation}
G=\frac{e^2}{h}\left[K_{\rm w}-
\frac{V_0^2 K_{\rm w}^2}{2\omega_c^2} \frac{\sqrt{\pi}\Gamma[K_{\rm w}]}{\Gamma[1/2+K_{\rm w}]}\left(\frac{\pi}{\beta\omega_c}\right)^{2K_{\rm w}-2} \right] \, .
\label{eq:infinite_Cond}
\end{equation}
However, note that this situation is unphysical, namely in order to generate a current through the interacting wire, we have to connect the system to the electronic leads, which forces us to take into account the finite-length of the wire and the resulting frequency structure of the Luttinger liquid parameter as we shall see in the following.
\subsubsection{Finite wire connected to Ohmic leads}
\begin{figure}[!htbp]
 \centering
\includegraphics[width=0.6\linewidth]{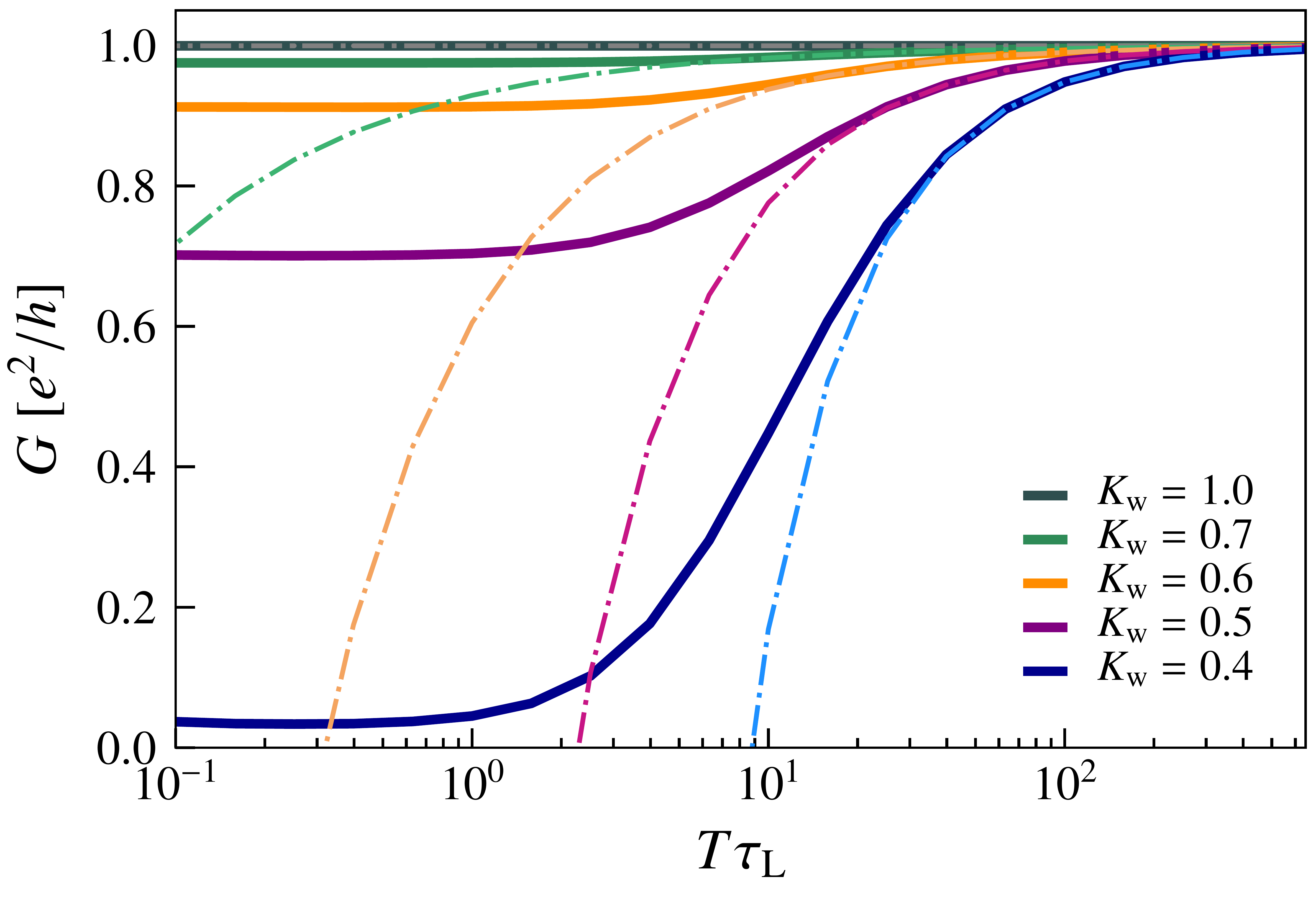}
 \caption{The conductance as a function of rescaled temperature for various interaction strengths ($K_{\rm w}\leq 1$), cf.~Eqs.~\eqref{eq:I_tilde} and~\eqref{eq:E_tau_simp}. The dashed-dotted lines correspond to Eq.~\eqref{con_ohmic_finite_L}.}
 \label{fig:con_ohmic}
\end{figure}

We now consider a finite-length wire, where the Luttinger Liquid parameter acquires a frequency dependence, as shown in the discussion related to Eq.~\eqref{eq:kLL_omega}. We evaluate the integrals [Eqs.~\eqref{eq:I_tilde} and~\eqref{eq:E_tau_simp}] numerically and show the results in Fig.~\ref{fig:con_ohmic}.
At high temperatures $T\tau_{L}>1$, the conductance can be approximated as  
\begin{equation}
G\approx \frac{e^2}{h}\left[1-
\frac{V_0^2}{2\omega_c^2} \frac{\sqrt{\pi}\Gamma[K_{\rm w}]}{\Gamma[1/2+K_{\rm w}]}\left(\frac{\pi}{\beta\Lambda}\right)^{2K_{\rm w}-2} \right],
\label{con_ohmic_finite_L}
\end{equation}
which has been obtained similar to the previous case of infinitely large wire [cf.~Eq.~\eqref{eq:infinite_Cond}, albeit considering that the dc-limit of the effective Luttinger liquid parameter is 1 (i.e., $K(\omega\to 0)=1$).
\subsection{Infinite wire connected to Ohmic-class leads}
We consider a generic Ohmic-class environment with the noise power spectrum defined in Eq.~(2) in the main text, and obtain the temperature dependence of the self-energy, and hence impurity-induced back-scattering conductance. We can rewrite Eq.~\eqref{eq:E_tau_simp}  as $E(it)=E_1(it)+E_2(it)$, with the first term being the temperature-independent part
\begin{align}
E_1(it)&=-\gamma^2K_{\rm w} (\omega_c)^{1-s}\int_{0}^{\infty} \frac{d\omega}{\pi}  
\omega^{s-2} e^{-\omega/\omega_c}
\left[1-e^{-i\omega t}\right] \nonumber\\
&=K_{\rm w} \frac{-\gamma^2}{\pi}\Gamma[s-1]
\left[1-\left(1+i\omega_c t\right)^{1-s}\right]\,,
\end{align}
and $E_2(\tau)$ being the temperature-dependant component
\begin{align}
\label{eq:E2_tau_general}
E_2(it)&=-(\omega_c)^{1-s}\gamma^2K_{\rm w}\int_{0}^{\infty} \frac{d\omega}{\pi}  
\omega^{s-2} e^{-\omega/\Lambda}
\left[\coth(\beta\omega/2)-1\right]
\left[1-\cos(\omega t)\right]\\
&=-\frac{\gamma^2}{\pi} \Gamma[s-1] \left(\beta \omega_c\right)^{1-s}
\Bigg\{
2\zeta\left[s-1,1+\frac{1}{\beta\omega_c}\right]
-\zeta\left[s-1,1+\frac{1}{\beta\omega_c}-\frac{it}{\beta}\right]
-\zeta\left[s-1,1+\frac{1}{\beta\omega_c}+\frac{it}{\beta}\right]
\Bigg\}\,,\nonumber
\end{align}
with $\zeta$ the Riemann zeta-function. The linear (in frequency) component of the self-energy [see Eq.~\eqref{eq:I_R}] then reads
\begin{align}
\tilde{I}&=-V_0^2(\beta/2)\int_{-\infty}^{\infty}
dt~e^{E_1(it)+E_2(it)} \nonumber \\
&=-V_0^2(\beta/2)~e^{-K_{\rm w}\alpha_s}
\int_{-\infty}^{\infty} dt~
e^{-K_{\rm w}\alpha_s\left\{
(1+i\omega_ct)^{1-s}+(\beta\omega_c)^{1-s}\left(
2\zeta[s-1,1+\frac{1}{\beta\omega_c}]
-\sum_{\eta=\pm}\zeta[s-1,1+\frac{1}{\beta\omega_c}+\eta\frac{it}{\beta}]
\right)
\right\}} \, ,
\label{eq:I_tilde_general}
\end{align}
with
$\alpha_s=\frac{\gamma^2}{\pi}\Gamma(s-1)$.

\subsubsection{sub-Ohmic case}

In the sub-Ohmic case, the current-current fluctuations are more pronounced at smaller frequencies, and hence at low-temperatures, the environmental effects become dominant. For this case $s<1$, the temperature dependant part, namely Eq.~\eqref{eq:E2_tau_general} is important. By changing the integration variable $t \to t+i\beta/2$ in Eq.~\eqref{eq:I_tilde_general} and using $\zeta[s-1,1+q]=\zeta[s-1,q]-q^{-(s-1)}$, we obtain
\begin{align}
\tilde{I}=-(\beta/2) V_0^2 e^{K_{\rm w}\alpha_s} \int dt  e^{
K_{\rm w}\alpha_s(\beta\omega_c)^{1-s}\left[
2\zeta[s-1,1+\frac{1}{\beta\omega_c}
-\sum_{\eta=\pm}\zeta[s-1,\frac{1}{\beta\omega_c}+\frac{1}{2}+\eta i \frac{t}{\beta}]
]
\right]} \, . 
\label{eq:tilde_I_sub}
\end{align}
As we are interested in the low-temperature behaviour,  $\omega_c \beta \gg 1$, at which the prefactor $(\beta\omega_c)^{1-s} \gg 1$, we can employ the method of steepest descend. In other words, we have
\begin{align}
\sum_{\eta=\pm}\zeta[s-1,\frac{1}{\beta\omega_c}+\frac{1}{2}+\eta i \frac{t}{\beta}]\approx 2\zeta[s-1,\frac{1}{\beta\omega_c}+\frac{1}{2}]
-\left(\frac{t}{\beta}\right)^2\zeta[s+1,\frac{1}{\beta\omega_c}+\frac{1}{2}]s(s-1),
\end{align}
which makes the integral in Eq.~\eqref{eq:tilde_I_sub} Gaussian, resulting in
\begin{align}
\tilde{I}=V_0^2 A_s e^{-B_s T^{1-s}} T^{\frac{s+3}{2}},
\end{align}
with
\begin{align}
&A_s=e^{K_{\rm w}\alpha_s}
\left(
\frac{\pi}{4K_{\rm w}\Gamma[s+1]\zeta[s+1,\frac{1}{2}+\frac{1}{\beta\omega_c}]\omega_c^{1-s}}
\right)^{1/2} \, , \nonumber\\
&B_s=2\omega_c^{1-s}\zeta[s-1,1+\frac{1}{\beta\omega_c}].
\label{eq:coefficients_sub_ohmic}
\end{align}
Since we are in the limit $\beta\omega_c\gg 1$, further temperature dependence in Eqs.~\eqref{eq:coefficients_sub_ohmic} can be neglected.
In conclusion, we found that the temperature dependence of the self-energy reads $\Sigma(i\omega)=\omega A_s \exp[-B_s T^{s-1}]T^{(s+3)/2}$, cf.~Fig.~(2) in the main text.

\subsubsection{super-Ohmic case}
We now turn into the super-Ohmic case $s>1$,  cf.~Fig.~(2) in the main text. The self-energy is
\begin{align}
\tilde{I}=-(\beta/2) V_0^2\lim_{\omega\to 0} \int dt~ e^{i\omega t} e^{E(it+\beta/2)}=
-(\beta/2) V_0^2 e^{K_{\rm w}\alpha_s}e^{-B_s T^{1-s}}
\int dt ~ e^{i\omega t}~e^{\tilde{E}(t)} \, ,
\end{align}
where we have defined 
\begin{align}
E(it+\beta/2)=\int_{0}^{\infty} d\omega~ \mathcal{G}_{\varphi\varphi}(i\omega)
\left\{\coth(\beta\omega/2)-\frac{\cos(\omega t)}{\sinh(\beta\omega/2)}\right\}\equiv \tilde{E}_0+\tilde{E}(t) .  
\end{align}
In contrast to the sub-ohmic case, in the low-temperature limit the pre-factor $(\beta\omega_c)^{1-s}\ll 1$, and hence we can expand the exponential in Eq.~\eqref{eq:tilde_I_sub} obtaining
\begin{align}
\tilde{I}\approx -(\beta/2) V_0^2 e^{K_{\rm w}\alpha_s} \frac{1}{2}\int dt e^{i\omega t} \tilde{E}^2(t) =
-(\beta/2) V_0^2 \int_{0}^{\infty} d\omega \frac{\mathcal{G}^2_{\varphi\varphi}(i\omega) }{\sinh^2(\beta\omega/2)}\propto T^{2s-4}.  
\end{align}
%
\end{document}